\documentclass[12pt]{iopart}
\usepackage{amssymb}
\usepackage{amstext}
\usepackage{color}
\usepackage{appendix}
\usepackage{graphicx}

\newcommand{\ket}[1]{\left\vert #1 \right\rangle}

\newcommand{\bra}[1]{\left\langle #1 \right\vert}

\newcommand{\eqref}[1]{(\ref{#1})}

\begin{document}
\title[On the existence of superradiant excitonic states in microtubules]{On the existence of superradiant excitonic states in microtubules}

\author{G.~L.~Celardo}
\address{$^1$ Benem\'erita Universidad Aut\'onoma de Puebla, Apartado Postal J-48, Instituto de F\'isica,  72570, Mexico}
\ead{celardo@ifuap.buap.mx}

\author{M. Angeli}
\address{$^2$ International School for Advanced Studies (SISSA), Via Bonomea 265, I-34136 Trieste, Italy}
\address{$^3$ Dipartimento di Matematica e
  Fisica and Interdisciplinary Laboratories for Advanced Materials Physics,
  Universit\`a Cattolica del Sacro Cuore, via Musei 41, I-25121 Brescia, Italy}
\ead{mangeli@sissa.it}

\author{P. Kurian$^*$}
\address{$^4$ Quantum Biology Laboratory, Howard University, Washington DC, USA 20059}
\address{$^5$ Center for Computational Biology and Bioinformatics, Howard University College of Medicine, Washington DC, USA 20059}
\ead{pkurian@howard.edu ($^*$corresponding author)}

\author{T. J. A. Craddock}
\address{$^6$ Departments of Psychology and Neuroscience, Computer Science, and Clinical Immunology, Nova Southeastern University, Fort Lauderdale FL, USA 33314}
\address{$^7$ Clinical Systems Biology Group, Institute for Neuro-Immune Medicine, Fort Lauderdale FL, USA 33314}
\ead{tcraddock@nova.edu}

\begin{abstract}
  Microtubules are biological protein polymers with critical and diverse functions. Their structures share some similarities with photosynthetic antenna complexes, particularly in the ordered arrangement of photoactive molecules with large transition dipole moments. As the role of photoexcitations in microtubules remains an open question, here we analyze tryptophan molecules, the amino acid building block of microtubules with the largest transition dipole strength. By taking their positions and dipole orientations from realistic models capable of reproducing tubulin experimental spectra, and using a  Hamiltonian widely employed in quantum optics to describe light-matter interactions, we show that such molecules arranged in their native microtubule configuration exhibit a superradiant ground state, which represents an excitation fully extended on the chromophore lattice.  We also show that such a superradiant ground state emerges due to supertransfer coupling between the ground states of smaller blocks of the microtubule. In the dynamics we find that the spreading of excitation is ballistic in the absence of external sources of disorder and strongly dependent on initial conditions. The velocity of photoexcitation spreading is shown to be enhanced by the supertransfer effect with respect to the velocity one would expect from the strength of the nearest-neighbor coupling between tryptophan molecules in the microtubule. Finally, such structures are shown to have an enhanced robustness to static disorder when compared to geometries that include only short-range interactions. These cooperative effects (superradiance and supertransfer) may induce ultra-efficient photoexcitation absorption and could enhance excitonic energy transfer in microtubules over long distances under physiological conditions. 
\end{abstract}

\pacs{71.35.-y, 87.16.Ka, 05.60.Gg, 87.15.bk, 87.15.hj}

\noindent{\it Keywords\/}: quantum biology; quantum transport in disordered systems; open quantum systems; energy transfer. 


\section{Introduction}

Since the discovery of coherent wave behavior stimulating
excitonic transport in natural photosynthetic
systems under ambient conditions ~\cite{photo,photoT},
ample motivation has arisen to investigate the relevance of quantum mechanical behavior in diverse biological networks of photoactive molecules. For instance, in photosynthetic systems, great attention has been
devoted to the antenna complexes. Such complexes are made of a network
of chlorophyll molecules (photoactive in the visible range), 
which are able to absorb sunlight and transport the excitation to a specific molecular aggregate (the
reaction center). The reaction center is where charge separation occurs, in order to trigger the ensuing steps required for carbon fixation~\cite{schulten}.  

Some of the dominant coherent effects which are thought to be responsible
for the high efficiency of natural photosynthetic complexes are
induced by the delocalization of the excitation over many 
molecules. Such delocalized excitonic states 
can lead to cooperative effects, such as superabsorption
and supertransfer~\cite{schulten,srlloyd}, and they can be useful 
in both natural and engineered 
light-harvesting complexes~\cite{kaplan,kaplan2,kaplan3,kaplan4,kaplan5,kaplan6,kaplan7,kaplan8,arvi1,sr2,vangrondelle,srfmo,srrc,K1,K2,K4}. 
Specifically, delocalized excitonic states can have a much 
larger dipole strength than that of the constituent chromophores, and such giant transient dipoles~\cite{fidder,mukamelspano,K3} can strongly couple to
the electromagnetic field. Thus, these states are able to
superabsorb light, i.e., they are able to absorb light at a rate which
is much larger than the single-molecule absorbing rate~\cite{K3}. 
Indeed, the absorption rate of delocalized excitonic states can
increase with the number of molecules over which the excitation is
delocalized~\cite{mukamelspano,K3}. Supertransfer is described in a similar way, with respect to movement of the excitation to an external molecular aggregate or between different parts of the same system~\cite{srlloyd}. Specifically, an excitonic state delocalized on $N$ molecules of one molecular aggregate  can couple with an excitonic state delocalized on $M$ molecules of a second aggregate with a coupling amplitude which is $\sqrt{NM}$ times larger than the coupling amplitude between single molecules belonging to different aggregates. Such supertransfer coupling  is able to enhance the velocity of spreading of photoexcitations, and  it has been shown to have an important role in natural photosynthetic systems~\cite{lh2}. 

The role of coherent energy transfer has been investigated not only in photosynthetic complexes but also  in 
other important biomolecular polymers, such as in cytoskeletal
microtubules~\cite{craddock, kurian}
and in DNA~\cite{JTB}. In this paper we will focus on the role of photoexcitations in
microtubules, which are essential biomolecular structures that have multiple
roles in the functionality of cells. Indeed, microtubules are
present in every eukaryotic cell to provide structural integrity to the cytoskeletal matrix, and they are
thought to be involved in many other cellular functions, including motor trafficking,
cellular transport, mitotic division, and cellular signaling in neurons. 
Interestingly, microtubules share some structural similarities with
photosynthetic antenna complexes, such as the cylindrical arrangement
of chlorophyll molecules in phycobilisome antennas~\cite{nir1} or in
green sulphur bacteria~\cite{alan}, where cylinders made of more than
$10^5$ chlorophyll molecules can efficiently harvest
sunlight for energy storage  in the form of
sugar. Note that while chlorophyll molecules are active in the visible range of electromagnetic radiation,  microtubules possess an architecture of chromophoric molecules (i.e., aromatic amino acids like tryptophan) which are photoactive in the ultraviolet (UV) range.

It remains an open question whether microtubules have any role
in transporting cellular photoexcitations. Intriguingly, several
groups have studied and experimentally confirmed the presence of very
weak endogenous photon emissions within the cell across the UV, visible, and IR
spectra~\cite{Kaznacheev, Quickenden, Slawinski, Scholkmann, Cifra}. It has also been suggested that microtubules may play a role in cellular orientation and ``vision" via the centrosome complex \cite{Albrecht-Buehler}, and very recently two of us have proposed neuronal signaling pathways in microtubules via coherent excitonic transport~\cite{kurian}. 

\begin{figure}
  \begin{indented}\item[]
    \centering
    \Large{a)} \includegraphics[scale=0.4]{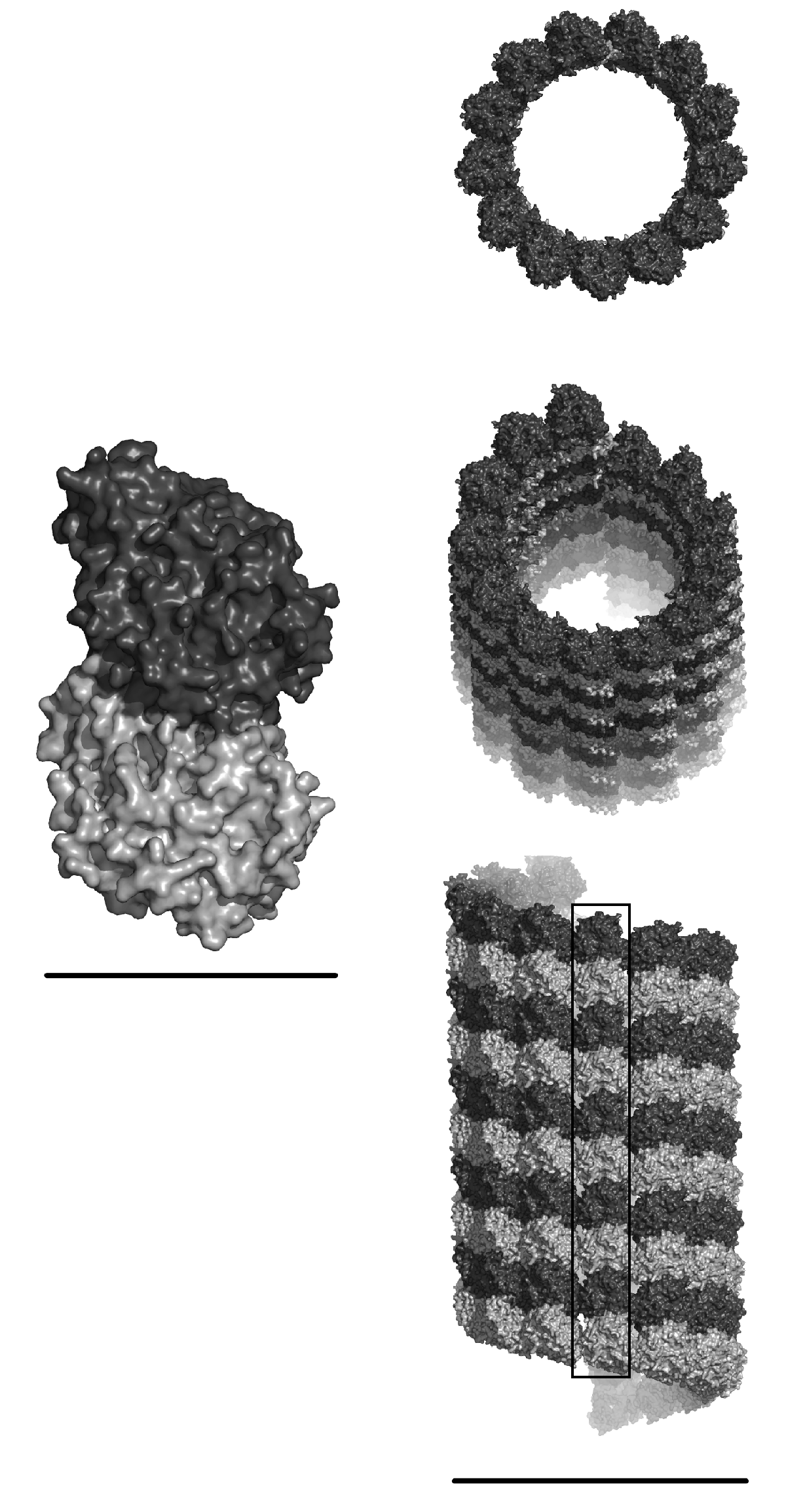}
  \Large{b)} \includegraphics[scale=0.1]{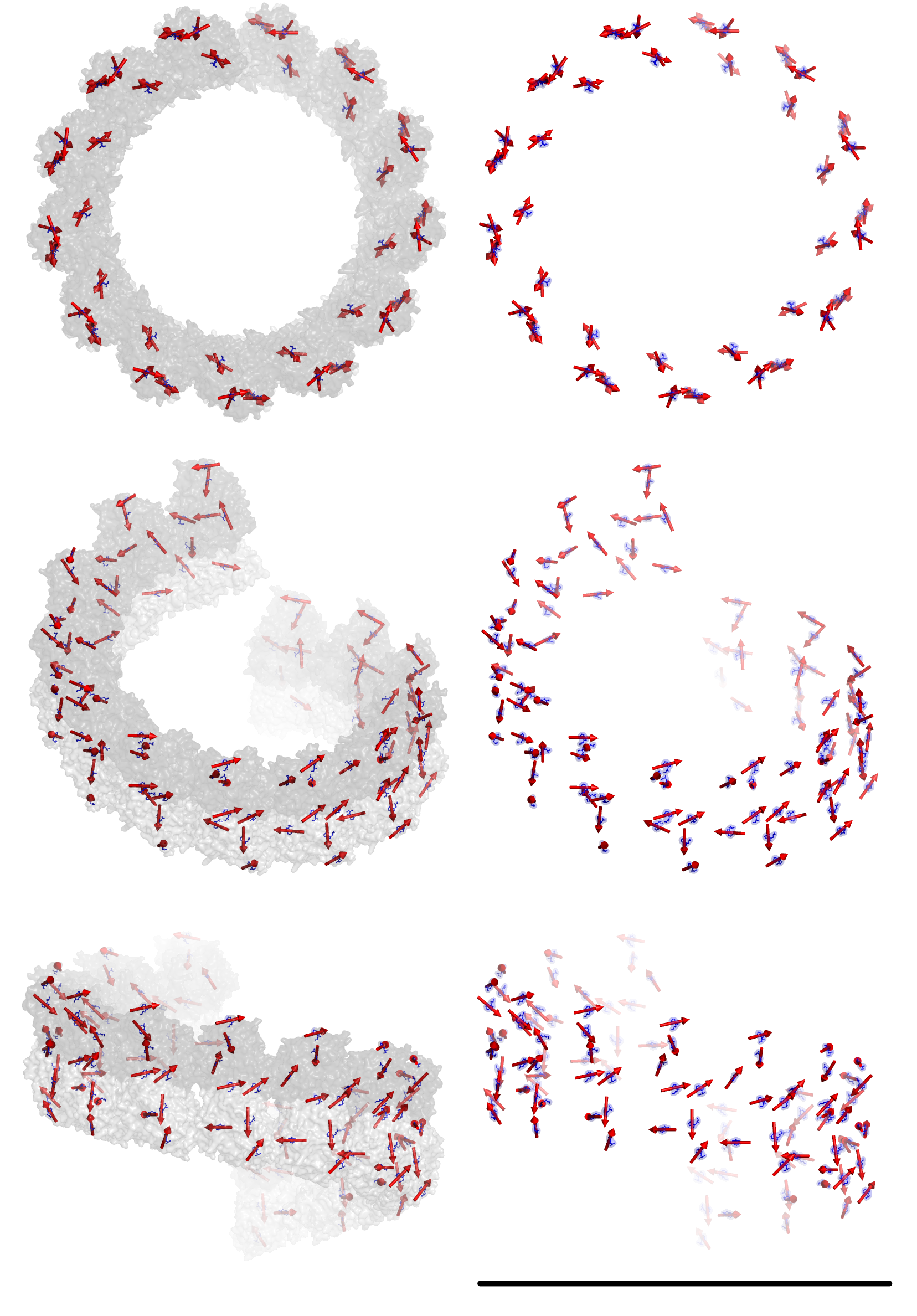}      
    \end{indented}
\caption{\\Panel a): Tubulin dimer and microtubule segment.  Left: Solvent-excluded tubulin heterodimer surface with $\alpha$-tubulin monomer in light grey, and $\beta$-tubulin monomer in dark grey (scale bar $\sim$5 nm).  Right: Section of microtubule B-lattice structure at three angles showing left-handed helical symmetry and protofilament (bottom, outlined in black box) (scale bar $\sim$25 nm).  In microtubules, tubulin dimers stack end-to-end to form the protofilaments, 13 of which join side-by-side with longitudinal offset and wrap around to form a tube with such helical symmetry. Panel b): Arrangement of tryptophan amino acids in microtubule segment at three angles with transition dipole directions.  Left: A single spiral of tubulin dimers (light grey $\alpha$-tubulin, dark grey $\beta$-tubulin) from microtubule structure showing tryptophan amino acids (blue sticks) and transition dipole directions (red arrows).  Right: Tryptophan amino acids and transition dipole directions only (scale bar $\sim$25 nm).}
\label{figMT}
\end{figure} 

Each mammalian microtubule is composed of 13 protofilaments, which form a helical-cylindrical arrangement of tubulin subunit protein dimers, as in Fig. (\ref{figMT}).
The tubulin subunit proteins possess a unique network of chromophores, 
namely different amino acids, which can
form excited-state transition dipoles in the presence of photons.
The geometry and dipole moments of these amino acids,
which are termed \textit{aromatic} owing to their largely delocalized $\pi$ electrons, are similar to
those of photosynthetic constituents, indicating that tubulin may
support coherent energy transfer. 
As with chlorophyll molecules, it is
possible to associate to each aromatic amino acid a transition dipole that determines its coupling to other molecules and with the
electromagnetic field. 

The main question we address in this paper is whether the arrangement of photoactive molecules in the microtubule structure can support extended
excitonic states with a giant dipole strength, at least in the absence of environmental disorder. Such extended states, if robust to noise, can also support efficient transport of photoexcitations, which could have a biological role in microtubule signaling between cells and across the brain~\cite{kurian}.
To answer this question, we consider first a quantum description of the network of tryptophan molecules, which are the greatest contributor to photoabsorption in microtubules in the UV range. 
Indeed, tryptophan is the aromatic amino acid with the largest transition dipole ($6.0$ debye), comparable with that of chlorophyll molecules.
We proceed by 
modelling these tryptophans as two-level systems, as is usually done in photosynthetic antenna complexes. This is, to our knowledge, the first analysis of excitonic states distributed across tryptophan chromophore lattices in large-scale, realistic models of microtubules.  

The interaction between the  transition dipoles of the photoactive molecules is in general very complicated, with the common coupling to the electromagnetic
field more nuanced than simple dipole-dipole interactions,
which are an effective description of a chromophoric network only in
the small-size limit (where the system size is much smaller then the wavelength)~\cite{haroche}. 
However, for large aggregates one needs to go beyond the simple dipole-dipole interactions used in small aggregates. Here we consider an effective non-Hermitian Hamiltonian interaction commonly used in the literature to study the coupling in molecular aggregates~\cite{mukamelspano}. 
This non-Hermitian description also allows the possibilities of donating the
excitation back to the electromagnetic field through photon emission or of transferring excitation
coherently between chromophores. Moreover, in the small-system-size limit it reduces to a dipole-dipole interaction. 

The imaginary part of the complex eigenvalues ${\cal E} = E
-i\Gamma/2$ of such a non-Hermitian Hamiltonian determines the strength of
the coupling of the excitonic states with the electromagnetic field and is connected with the dipole strength of the eigenstates of the system.
While the coupling of a single aromatic
molecule can be characterized by its decay rate $\gamma$, $\Gamma$
determines the coupling of extended excitonic states with the electromagnetic field. 
Superradiant states are characterized by $\Gamma>\gamma$, while
subradiant states are characterized by  $\Gamma < \gamma$. Most
importantly, for superradiant states, $\Gamma$ should be proportional to
the number of chromophores $N$, for $L \leq \lambda$, where $L$ is the system length
and $\lambda$ is the
wavelength of the aromatic molecule optical transition, and begins to saturate for $L > \lambda$. 
Note that   $\hbar/\Gamma$ is the lifetime of the excitonic
eigenstate, so that larger values of $\Gamma$ govern faster excitation
decays. Since the process is symmetric under time
reversal, fast decaying states are also fast absorbing states. 
The advantage of this formalism, with respect to the simple dipole-dipole interaction commonly used in the literature, is that it allows us to consider system
sizes that are even larger than the wavelength of the absorbed light.
This property becomes particularly important for large biopolymeric structures like
microtubules whose length is generally several orders of magnitude larger than the wavelength associated with the molecular transitions ($\lambda=280$ nm). Here we consider microtube lengths up to $\sim 3\lambda$. 

We use data on the positions, dipole orientations, and
excitation energies of tryptophan molecules, which have been obtained by
molecular dynamics simulations and quantum chemistry calculations \cite{craddock, kurian}. These data have been shown to reproduce well the
absorption, circular dichroism, and linear dichroism spectra of single tubulin dimers \cite{craddock, kurian}. 

Our analysis shows that as the number of tubulin subunits considered
grows, a superradiant state forms in the excitonic ground
state of the system. 
This is exactly what happens in many photosynthetic antenna complexes,
such as in green sulfur bacteria cylindrical antennas~\cite{K2,K4,gulli} and in self-assembled molecular nanotubes~\cite{nanotube,caoNT,eisele2,eisele3,JYZ,JRC}. Superradiant states favor the
absorption of photons by the microtubule. Moreover, since the superradiant ground state represents  an
extended (delocalized) excitonic state of the order of the microtubule length, such superradiant ground states  
could serve as a support for efficient transport of photoexcitation. 

In the next section we develop the mathematical machinery for our physical model, using an effective Hamiltonian that has been widely used to describe a single photoexcitation interacting within a network of transient dipoles. In Section 3 we display several  results demonstrating the existence of a superradiant excitonic ground state extended over more than $10^4$ tryptophan molecules of the microtubule. We also show that the superradiant ground state emerges from the supertransfer coupling between the superradiant ground states of smaller segments inside the microtubule. Section 4 shows initial studies of the exciton  dynamics, showing that cooperativity can enhance the coupling between different parts of the microtubule through  supertransfer. Section 5 demonstrates the robustness of the superradiant ground state to disorder, and we close in Section 6 with some conclusions and our future outlook.

\section{The model}

Microtubules are cylindrical-helical structures made of essentially two closely related proteins,
$\alpha$- and $\beta$-tubulin. They are arranged  as in Fig.~(\ref{figMT}a) to form a 
left-handed helical tube of protofilament strands. In each $\alpha-\beta$ dimer there are many aromatic molecules: eight tryptophans (Trps) whose transition dipoles are arranged as in Fig.~(\ref{figMT}b) (see \ref{app-a} for complete description). Their peak excitation
energy is $\sim280$nm, and the magnitude of their dipole moment is $6.0$ debye. There also exist  other aromatics, including tyrosine, phenylalanine, and histidine, with much smaller
dipole moments. For example, tyrosine (the molecule with the second largest dipole) has a dipole moment of only 1.2
debye. For the purposes of this initial analysis, we limit our
attention to the Trps only because of their relatively large
transition dipoles.  The position and orientation of the dipole
moments of Trp molecules have been obtained from molecular dynamics and quantum chemistry calculations, and
they reproduce closely the linear and circular dichroism spectra of tubulin for the Trp-only case~\cite{craddock,kurian}.

The interaction of a network of dipoles with the electromagnetic field
is well described by the effective
Hamiltonian~\cite{mukamelspano,mukameldeph,kaiser}
\begin{equation}
  H_{\it eff} = H_0 + \Delta - \frac{i}{2} \Gamma\, , \label{Hmuk}
\end{equation}
where $H_0$ represents the sum of the excitation energies of each molecule, and
$\Delta$ and $\Gamma$ represent the coupling between the molecules
induced by the interaction with the electromagnetic field.  Note
that such an effective Hamiltonian has been widely used to model
light-matter interactions in the approximation of a single excitation. 
The site energies are all identical, so that we have
\begin{equation}
  H_0 = \sum_{n=1}^N \hbar \omega_0 \ket{n} \bra{n}.
\end{equation}
The wavenumber associated with each site energy is $k_0:=\omega_0
n_r/c$, where $c$ is the speed of light and $n_r = \sqrt{\mu_r \epsilon_r}$ is the refractive
index. Most natural materials are non-magnetic at optical frequencies (relative permeability $\mu_r \approx 1$) so we can assume that $n_r \sim \sqrt{\epsilon_r}$. The real and imaginary parts of the intermolecular coupling are given on the diagonal, respectively, by
\begin{equation}
  \Delta_{nn} = 0 \, , \\
  \Gamma_{nn} = \frac{4}{3} \frac{\mu^2}{\epsilon_r} k_0^3 =: \gamma \, , \label{eq:gamma}
\end{equation}
with $\mu=|\vec{\mu}|$ being the magnitude of the transition dipole of a single tryptophan and $\epsilon_r$ being the relative permittivity, and on the off-diagonal, respectively, by
$$
  \Delta_{nm} = \frac{3\gamma}{4} \left[ \left( -\frac{\cos (k_0 r_{nm})}{(k_0 r_{nm})} +
    \frac{\sin (k_0 r_{nm})}{(k_0 r_{nm})^2} + \frac{\cos (k_0 r_{nm})}{(k_0 r_{nm})^3} \right)
    \hat{\mu}_n \cdot \hat{\mu}_m +\right. \nonumber \\
$$    
\begin{equation}
    -\left. \left( -\frac{\cos (k_0 r_{nm})}{(k_0 r_{nm})} + 3\frac{\sin (k_0 r_{nm})}{(k_0 r_{nm})^2} +
    3\frac{\cos (k_0 r_{nm})}{(k_0 r_{nm})^3}\right) \left( \hat{\mu}_n \cdot \hat{r}_{nm}
    \right) \left( \hat{\mu}_m \cdot \hat{r}_{nm} \right) \right],\\
    \label{eq:d1}
\end{equation}

$$  
  \Gamma_{nm} = \frac{3\gamma}{2} \left[ \left( \frac{\sin (k_0 r_{nm})}{(k_0 r_{nm})} +
    \frac{\cos (k_0 r_{nm})}{(k_0 r_{nm})^2} - \frac{\sin (k_0 r_{nm})}{(k_0 r_{nm})^3} \right)
    \hat{\mu}_n \cdot \hat{\mu}_m +\right. \nonumber \\
    $$
\begin{equation}    
  -\left. \left( \frac{\sin (k_0 r_{nm})}{(k_0 r_{nm})} + 3\frac{\cos (k_0 r_{nm})}{(k_0 r_{nm})^2} -
    3\frac{\sin (k_0 r_{nm})}{(k_0 r_{nm})^3}\right) \left( \hat{\mu}_n \cdot \hat{r}_{nm}
    \right) \left( \hat{\mu}_m \cdot \hat{r}_{nm} \right) \right], 
    \label{eq:g1}
\end{equation}
where $\hat{\mu}_n :=  \vec{\mu}_n  /
\mu$ is the unit dipole moment of the $n$th site and $\hat{r}_{nm} := \vec{r}_{nm}
/ r_{nm}$ is the unit vector joining the $n$th and the $m$th sites. In the following we assume $\epsilon_r = 1 = n_r$, corresponding to their values in vacuum/air. The actual dielectric constant and refractive index of tubulin is currently debated \cite{kurian,Cifra}, but using the tubulin dielectric instead of air would increase the imaginary part of the coupling by $\sqrt{2} \leq \sqrt{\epsilon_r} \leq \sqrt{8.41}$, depending on the value chosen, which would proportionally decrease the lifetimes of the excitonic eigenstates. The real part of the off-diagonal coupling would also be increased by the same factor, augmenting the dipole strength of the excitonic state.

The eigenvalues of this Hamiltonian are complex, endowing the eigenstates with a
finite lifetime due to their coupling to the external environment.
The imaginary part of an eigenvalue is directly linked to the decay
rate $\Gamma$ of the eigenstate. Thus for each eigenmode of the
system, $\Gamma$ represents its coupling to the electromagnetic field. 
For $\Gamma > \gamma$ we have excitonic states which are coupled to
the field more strongly than the single constituent molecule, representing
superradiant states. On the other hand, the states for which $\Gamma < \gamma$ are subradiant. 
Since the sum of all the widths of the excitonic states of a system must be  equal to $N \gamma$, where $N$ is the total number of chromophores, superradiant states are always found in conjunction with subradiant states. 
Note that in a large ensemble of molecules, a certain degree
of symmetry is needed to manifest superradiant and subradiant states~\cite{haroche}. 
In fact, a disordered network of dipoles suppresses superradiance, as we show below. 

The non-Hermitian character of the Hamiltonian  in Eq.~(\ref{Hmuk}) is due to the fact that the ensemble of photoactive molecules is not a closed system, since it interacts with the continuum of the electromagnetic field where the excitation can be radiatively lost. The analysis of open quantum systems within the framework of non-Hermitian Hamitonians is well-developed ~\cite{kaplan,kaplan2,fesh, zele1} and used in the field of quantum optics with applications to photosynthetic complexes~\cite{mukamelspano}. 
In this framework, the eigenstates of the non-Hermitian Hamiltonian represent the projection on the single excitation manifold of the true eigenstates of the molecular aggregate including also the photon degrees of freedom. 
So, if we indicate as $|{ \cal E}\rangle$ the eigenstate of the non-Hermitian Hamiltonian,
\begin{equation}
P(k)=\frac{|\langle k|{\cal E} \rangle|^2}{\sum_k |\langle k|{\cal E} \rangle|^2}
\label{pk}
\end{equation}
represents the conditional probability to find the excitation on site $k$ given that the excitation is in the system and not in the photon field. 
The time evolution of an initial state $|\psi_0 \rangle$ can be computed as 
$$
| \psi(t) \rangle= e^{-i H_{eff} t/ \hbar} | \psi_0 \rangle,
$$
and $| \psi(t) \rangle$ represents the projection on the single excitation manifold of the molecular aggregate full wave function (including also the photon field degrees of freedom) at time $t$. In order to compute such time evolution one has to consider that right $|{\cal E}^R \rangle$ and left $|{\cal E}^L \rangle$ eigenstates for a symmetric non-Hermitian Hamiltonian are the transpose of each other and not the Hermitian conjugate of each other. Moreover they represent a complete bi-orthogonal basis set: $ \langle {\cal E}^L_l| {\cal E}^R_m \rangle = \delta_{lm}$, which implies that the right eigenvectors  are normalized such that $\sum_k (\langle k| {\cal E}^R \rangle)^2=1$. 
Given an initial state $|\psi_0\rangle$, its 
decomposition  reads: $|\psi_0 \rangle = \sum_k \langle {\cal E}^L_k| \psi_0 \rangle |{\cal E}^R_k \rangle$. Note that from such a decomposition we can easily compute the time evolution of any initial state.  We will make use of the above considerations in the following sections.

The parameters considered in our analysis are~\cite{craddock, kurian}:
\begin{itemize}
\item $ e_{0} = 280 \text{ nm} = 35716.65$ cm$^{-1}$ as the Trp excitation energy, 
\item $ k_0 = 2 \pi e_0 \times 10^{-8} = 2.24 \times 10^{-3}$ \AA$^{-1}$ as the angular wavenumber,  
\item $\mu=6$ D as the strength of the transition dipole between the ground state and the first excited state, with $\mu^2 \approx 181224 \mbox{  \AA}^3 \mbox{  cm}^{-1}$ (for the conversion, see~\cite{dipsquare}),  
\item $ \gamma= 4 \mu^2 k_0^3 / 3 = 2.73 \times 10^{-3} \text{ cm}^{-1}$, where $ \gamma/\hbar $ is the radiative decay rate of a single Trp molecule, corresponding to the radiative lifetime $\tau_\gamma \approx 1.9 \text{ ns}$ (for the conversion, see~\cite{enertime}), and
\item  $n_s$=104 as the number of dipoles per microtubule spiral.
\end{itemize}

 It should be noted that for small systems $(k_{0}r_{ij}\ll 1)$ the coupling terms in the Hamiltonian (\ref{Hmuk}) become 
\begin{equation}
\begin{array}{lll}
\label{real}
\Gamma_{ij}&\simeq\displaystyle \gamma \hat{\mu}_i \cdot \hat{\mu}_j \,,\\
&\\
\Delta_{ij} &\simeq\displaystyle \frac{\vec{\mu}_{i} \cdot \vec{\mu}_{j}-3(\vec{\mu}_{i} \cdot \hat{r}_{ij})(\vec{\mu}_{j} \cdot \hat{r}_{ij})}{r_{ij}^{3}}.\\
\end{array}
\end{equation}
Here we have neglected terms that go as $1/r_{ij}$ because they are dominated by $1/r_{ij}^3$ contributions. In this limit, the real part $\Delta_{ij}$  represents a dipole-dipole interaction energy with $\mu =|\vec{\mu}_j|$ and the radiative decay width $\gamma=\frac{4}{3}\mu^{2}k_{0}^{3}$. 
Recall that the usual dipole-dipole coupling which is used to describe the
interactions between the transition dipoles of photoactive molecules
cannot be used in our case, since this approximation is valid only
when the size of the system $L$ is much smaller then the wavelength $\lambda$
associated with the transient dipole. In our analysis we
consider microtubule lengths which are larger than $\lambda$. 
Thus in our case it is mandatory to go beyond the dipole-dipole approximation (see discussion in \ref{app-b}).

\section{Superradiance in the ground state}

We have diagonalized the full radiative Hamiltonian given in Eq. (\ref{Hmuk}) 
for microtubule segments of different sizes, up to a microtubule
more than $800$ nm long and comprised of 100 spirals, including a total of $10400$ Trp molecules, so that $L/\lambda \approx 3$. 
For each eigenstate and complex eigenvalue ${\cal E}_k=E_k-i \Gamma_k/2$,
we plot the decay width $\Gamma_k$ of each state
normalized to the single dipole decay width $\gamma= 2.73 \times 10^{-3} \text{ cm}^{-1}$.

\begin{figure}
  \centering
\includegraphics[ trim=1cm 0 12 0,scale=0.6]{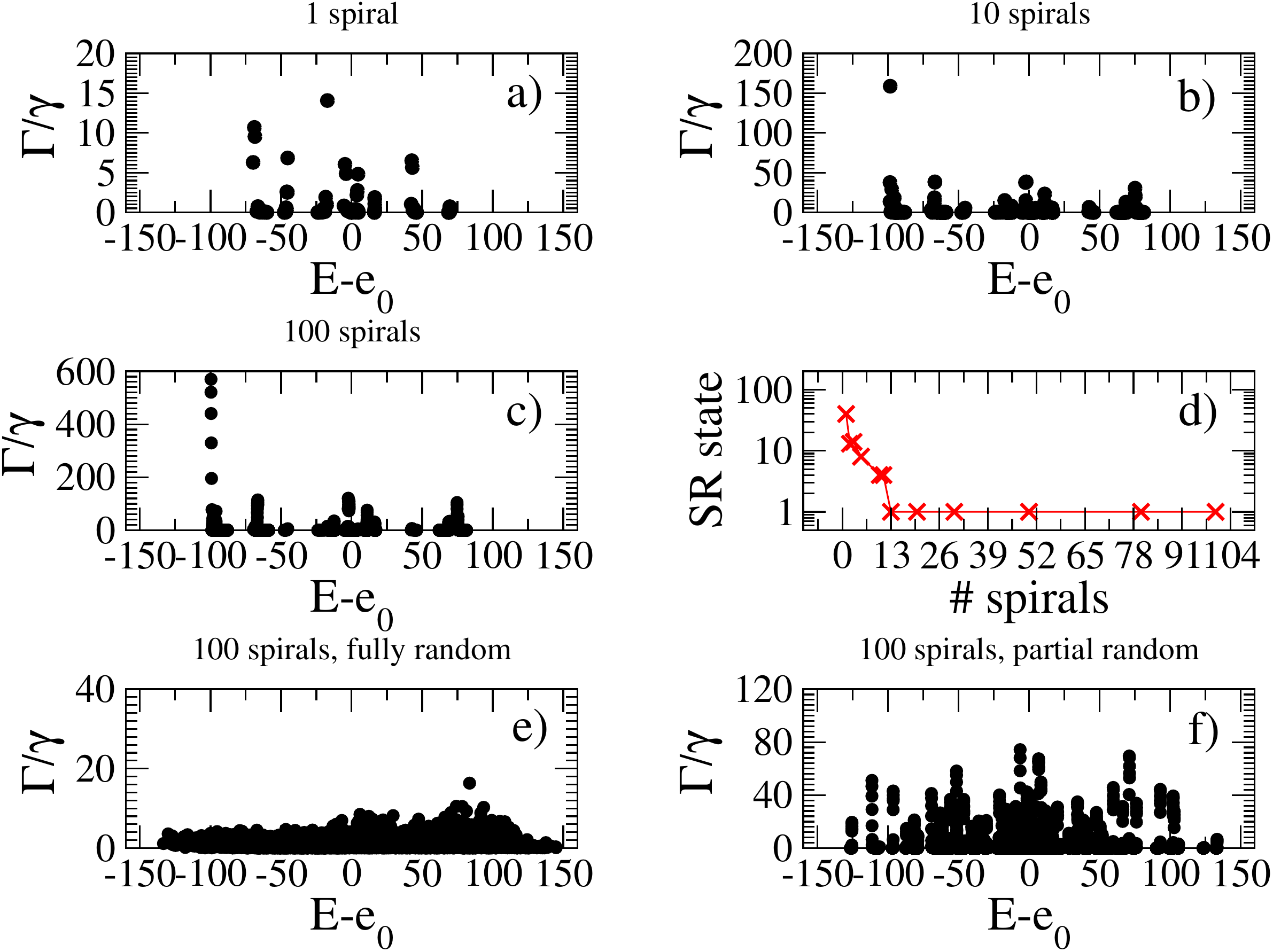}
\caption{ \\ Panels a,b,c,e,f: Normalized decay widths $\Gamma/\gamma$ of the excitonic eigenstates are plotted vs. their energies for microtubule segments of different lengths (number of spirals). Note that
  each spiral contains $104$ tryptophan molecules and extends about $9$ nm in the longitudinal direction. 
  In panel c) the maximum length of the microtubule segments considered in this paper is
  shown, comprised of 100 spirals and  10400 tryptophan molecules. 
  In panel e) a microtubule of the same length (100 spirals) is shown; the positions of the tryptophans are the same as in panel c), but the orientations of their
  dipoles are randomized. In panel f) a microtubule of 100 spirals is shown; the positions of the tryptophans are the same as in panel c), but the orientations of their
  dipoles are randomized in only one spiral and then repeated in all the other spirals. Finally in panel d) the location in the energy spectrum of the superradiant state is shown as a function of the number of spirals. The superradiant state coincides with the ground state (state 1) for all microtubule segments of length $> 12$ spirals.}
\label{f4}
\end{figure}

In Fig. (\ref{f4}) we show how cooperativity (superradiance) emerges as we increase the
number of spirals in the microtubule segment (where each spiral contains
$104$ Trp molecules). For one spiral (Fig. (\ref{f4}a)), there is a very
disordered distribution of the decay widths, but as we increase the number of spirals
a superradiant ground state clearly emerges with a decay width  $\Gamma_k > \gamma$ that increases
as we increase the length of the microtubule segment.  
In Fig. (\ref{f4}c) the normalized decay widths $\Gamma/\gamma$ vs. energies of the eigenstates of the
microtubule comprised of 100 spirals are shown. As one can see, most of
the decay width is concentrated in the ground state. The lowest-energy (ground) superradiant state in Fig.~(\ref{f4}c) corresponds to $\sim 600$ times the single-molecule decay rate. In Fig. (\ref{f4}d) the location in the energy spectrum of the largest superradiant state is shown as a function of the number of spirals. The energy of the largest superradiant state is indicated by an integer, where one means that the superradiant state is in the ground state, two that it is in the first excited state, etc. As one can see for all microtubule segments with number of spirals $> 12$, the superradiant state is in the ground state. 
The large decay width of the  superradiant ground state
indicates that such structures could be able to absorb photons ultra-efficiently.
Indeed, the decay width of the ground state
of the microtubule is in this case almost 600 times larger than the
single molecule decay width, corresponding to a value of roughly
$1.64 \text{ cm}^{-1}$. This translates to an absorbing time scale of $\hbar/\Gamma \approx 3.2$ ps, 
which is very fast and comparable with the typical thermal relaxation times for biological structures~\cite{schulten1}, suggesting that non-equilibrium processes might be relevant in this regime. 

\begin{figure}
    \centering
\includegraphics[ trim=1cm 0 12 0,scale=0.6]{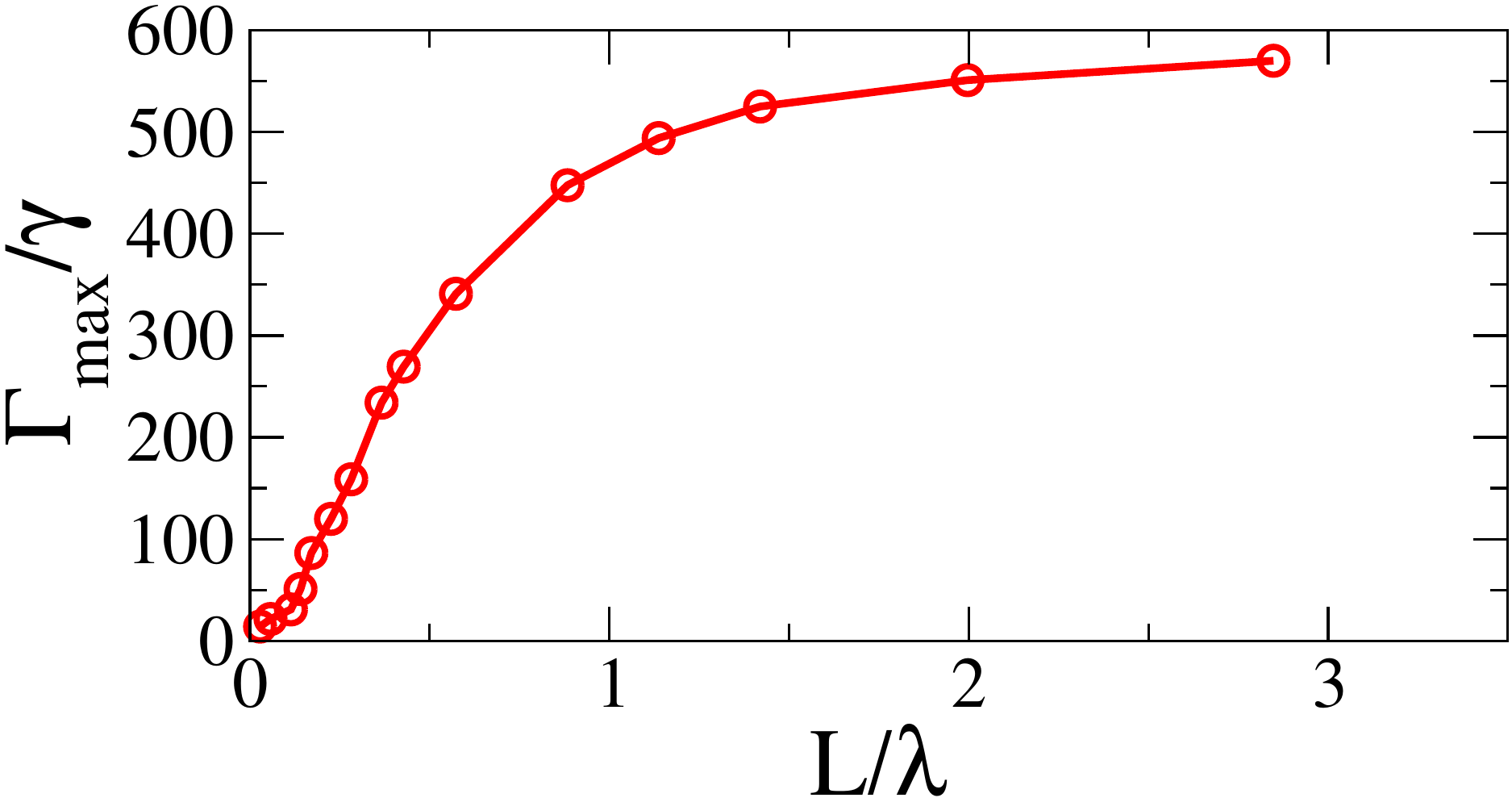}
\caption{ \\The maximum normalized decay width of the microtubule
  $\Gamma_{\rm max} /\gamma$ is shown as a function of the microtubule length $L$ rescaled by the wavelength of the light that excites the atoms, $\lambda=280$ $\rm nm$. } 
\label{f3}
\end{figure}

The existence of a superradiant ground state is surprising considering that the
positions and orientations of the dipoles may look quite disordered at first
sight, as shown in Fig.~({\ref{figMT}b}). It is well known that interactions between molecules can destroy superradiance~\cite{haroche} unless dipole orientations have a certain degree of symmetry. The orientations of the Trp dipoles in the microtubule are far from being random, and their symmetry plays an
important role. To show this we consider two additional models where the positions of the dipoles are the same as in the realistic case but with their orientations randomized. First we consider the case where the orientations of the dipoles are fully randomized over the whole microtubule length. In such a case the superradiance 
is completely suppressed, as shown in Fig.~(\ref{f4}e). Note also that in
Fig.~(\ref{f4}e) the decay widths are distributed
over many states, in contrast to the case of the native orientations of the dipoles,
where most of the decay width of the system is concentrated in the ground state. The maximum value of the decay width for randomized dipoles in 100 spirals is much smaller than that of the superradiant ground state shown in Fig.~(\ref{f4}c), and even smaller than some decay widths shown in Fig.~(\ref{f4}a) for one spiral. One might also think that the emergence of a superradiant ground state is connected with the fact that the same dipole geometry is repeated over all the spirals. To show that this is not the case, we considered a second random model with random orientations of dipoles on a single spiral repeated over all other spirals. For this partial random model we still do not achieve a superradiant ground state, as shown in Fig.~(\ref{f4}f).

\begin{figure}
    \centering
 \includegraphics[ trim=1cm 0 12 0,scale=0.6]{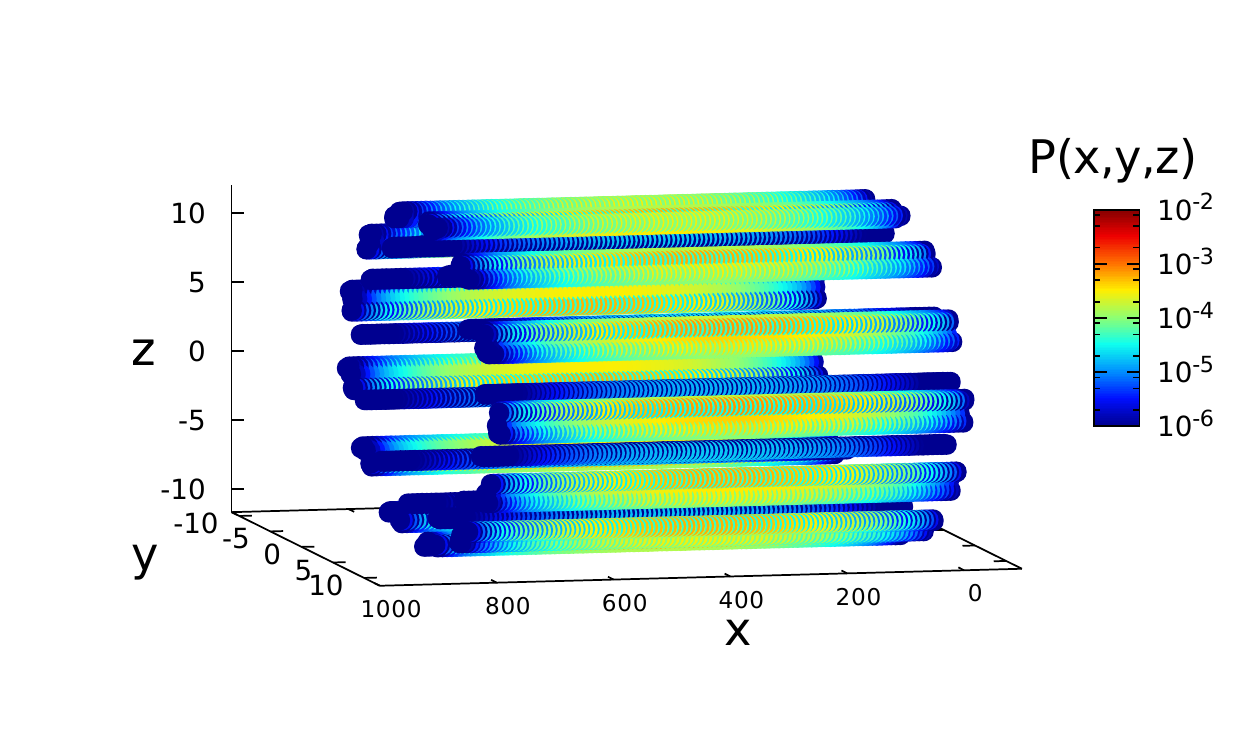}
\includegraphics[ trim=1cm 0 12 0,scale=0.6]{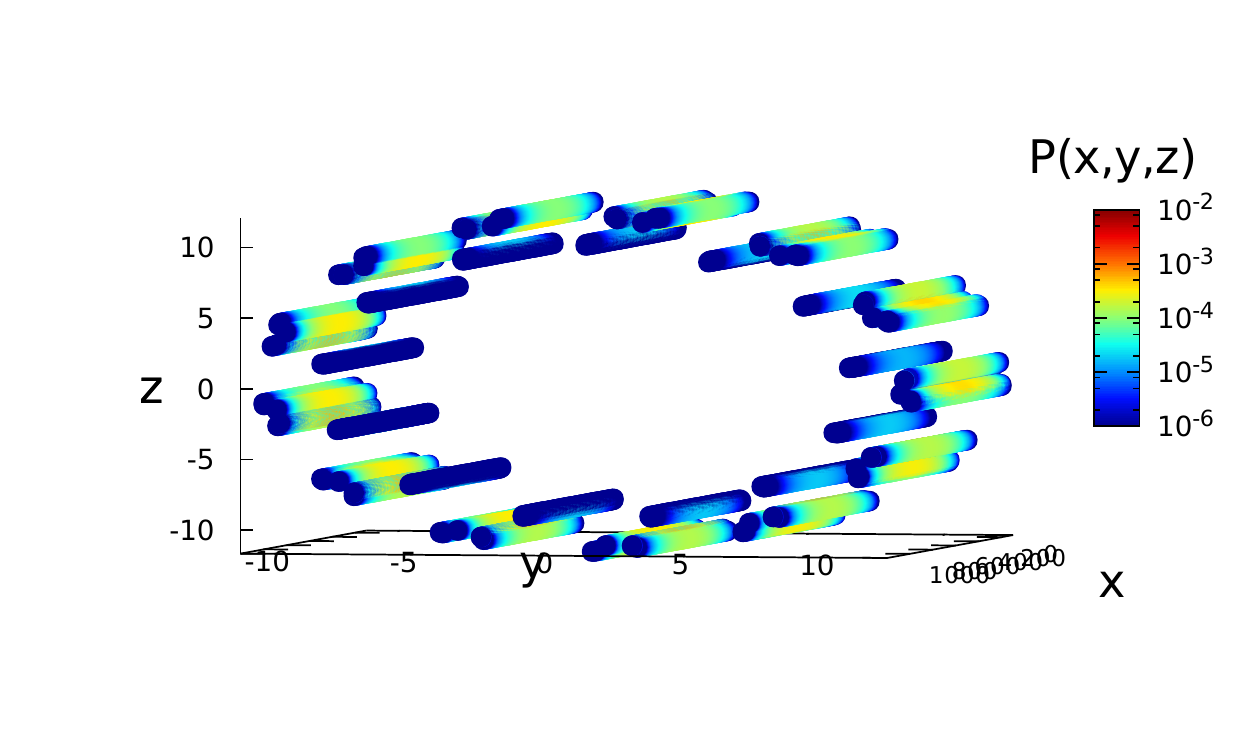}
\includegraphics[ trim=1cm 0 12 0,scale=0.6]{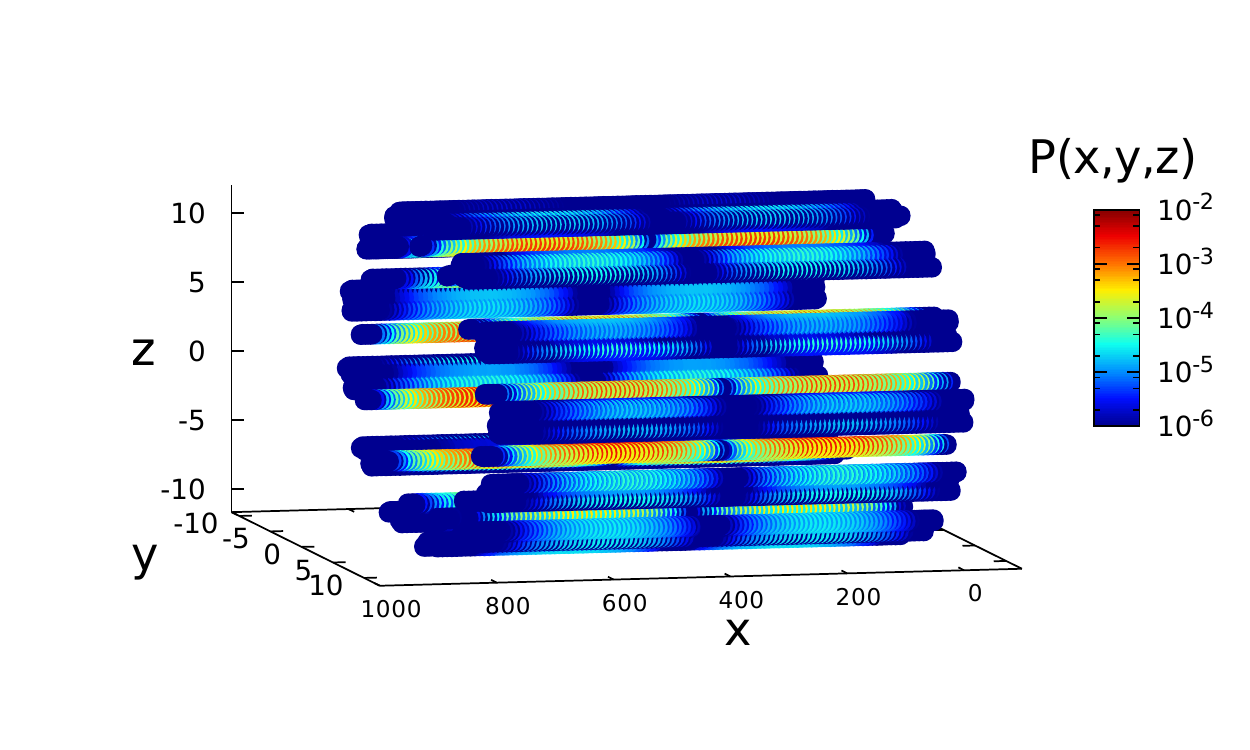}
\includegraphics[ trim=1cm 0 12 0,scale=0.6]{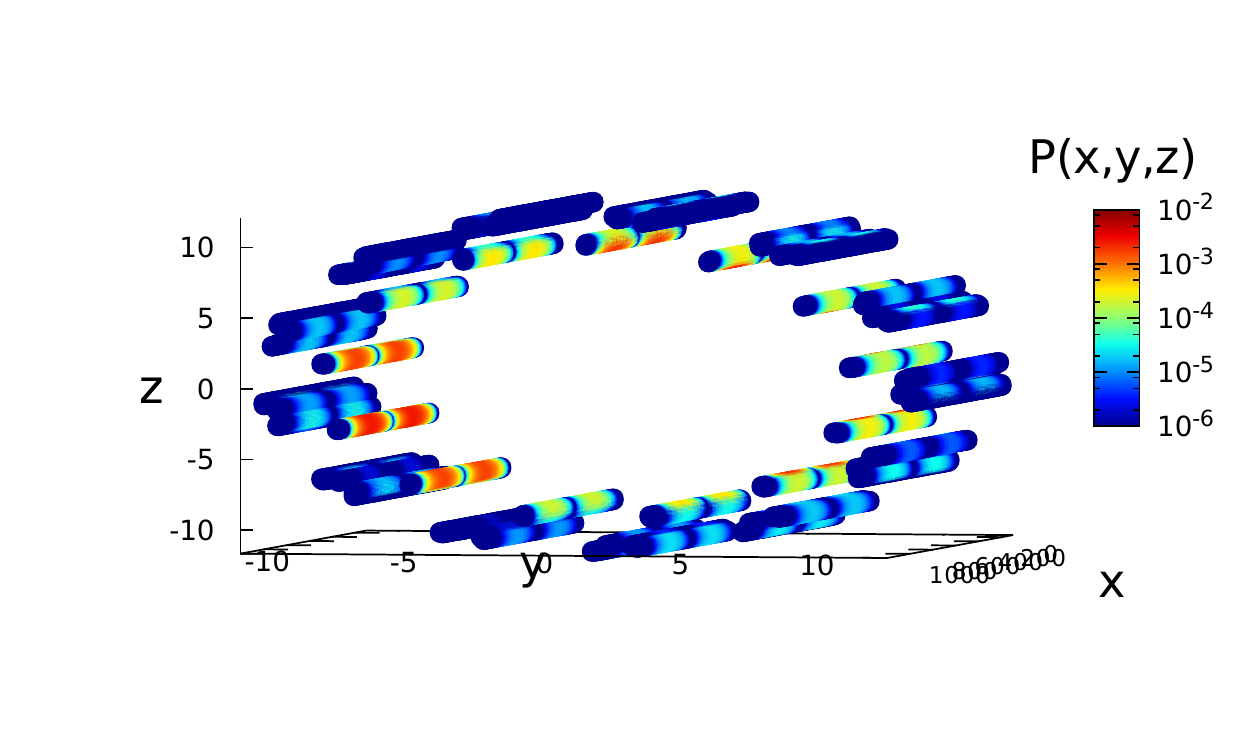}
\caption{ {\it Superradiant and subradiant excitonic states in a microtubule.}\\ The probability $P(x,y,z)=|\psi|^2$ of finding the exciton on a tryptophan chromophore of a microtubule segment of 100 spirals is shown for the extended superradiant ground state (upper panels, lateral view (left) and in cross section (right)) and the most subradiant state (lower panels, lateral view (left) and in cross section (right)), which has the smallest decay width ($\Gamma / \gamma \simeq 10^{-8}$) and an energy in the middle of the spectrum ($E-e_0=2.1 \text{ cm}^{-1}$). Lengths on each axis are expressed in nanometers.}
\label{sup}
\end{figure} 

In order to understand how the superradiant decay width increases
with the system size, 
in Fig.~(\ref{f3}) the maximum decay width is plotted as a function of
the length of the microtubule segment. Note that the decay width increases
with the system size, but saturation occurs when the
length of the microtubule is larger than $\lambda$, the wavelength
associated with absorption by the transient dipoles. 

Such a large decay width of the superradiant ground state indicates that
the excitation in the ground state is extended
over many Trp molecules. In the upper panels of Fig.~(\ref{sup}),
the probability of finding the excitation on each Trp molecule (see Eq.~(\ref{pk})) is shown when the system of 100 spirals is in its ground state. One can see that the ground
state represents a fully extended excitonic state over the whole
microtubule segment, and thus it could be capable of supporting ultra-efficient transport of photoexcitation.  
Note that the superradiant ground state is the state which is most strongly coupled to the electromagnetic field (highest decay width), and thus the fact that it represents an extended state implies that the absorbed photon will be shared by many tryptophan molecules in a coherent way, at least up to the dephasing time. In the lower panels of Fig.~(\ref{sup}), we also show for comparison the most subradiant state for a microtubule of 100 spirals. Note that in this case the excitation probability is concentrated on the chromophores of the inner wall of the microtubule lumen, contrary to the superradiant state where the excitation probability is delocalized on the chromophores of the external wall that forms an interface with the cytoplasm. 

\subsection{Structure of the superradiant ground state, super and subtransfer processes}
\begin{figure}
    \centering
\includegraphics[ trim=1cm 0 12 0,scale=0.6]{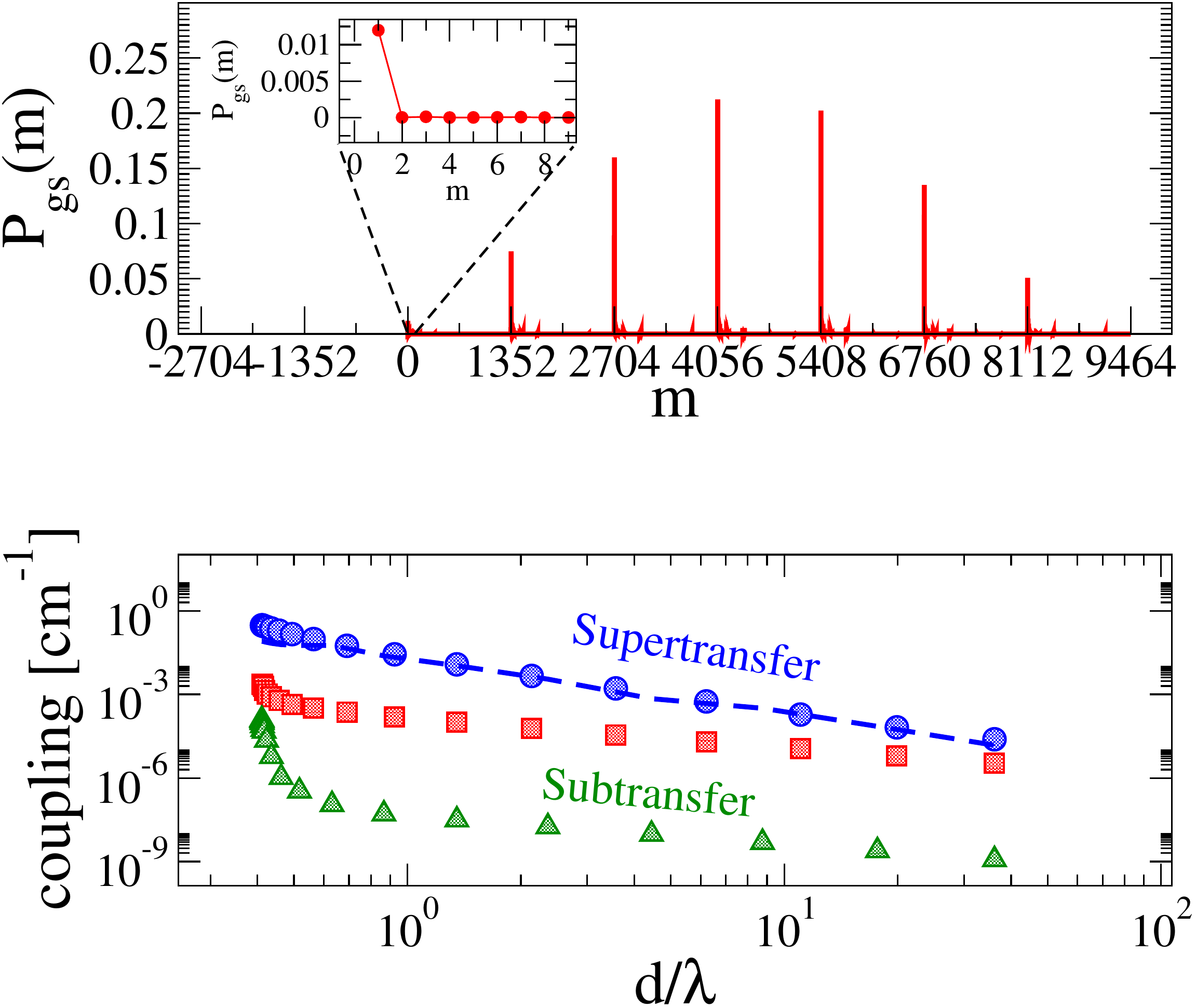}
\caption{\\ Upper panel:  Projection  of the ground state $|\psi_{gs} \rangle$ of a microtubule segment of 91 spirals over the basis states $|\phi_m \rangle$ built from the eigenstates of blocks of 13 spirals within the segment (see text). The basis states, indexed by $m$, are ordered from low to high energy in each block (13 $\times$ 104 = 1352 states). The projection has been computed as $P_{gs} (m)= |\langle \phi_m | \psi_{gs} \rangle|^2/\sum_m |\langle \phi_m | \psi_{gs} \rangle|^2$. The inset, zooming in on the first nine eigenstates of the first block, confirms that the ground state of the whole microtubule segment can be viewed as a coherent superposition of the ground states of the smaller blocks of spirals. Lower panel: Coupling between the superradiant ground states of two blocks of 13 spirals (blue circles) is compared with the average pairwise coupling between the chromophores of each block (red squares) and the most subradiant states of the two blocks (green triangles). The couplings are plotted versus the distance $d/\lambda$ (normalized by the excitation wavelength $\lambda=280$ nm) between the centers of the two blocks. When two blocks are immediate neighbors, the center-to-center distance is $d \approx 116$ nm. The supertransfer interaction between the giant dipoles of the ground states of the two blocks (see Eq.~(\ref{eq:st2})), valid for large inter-block distances $d$, is shown as a blue dashed curve.} 
\label{Proj}
\end{figure}

In order to understand the structure of the superradiant ground state of
a large microtubule segment, we now project the ground state of the whole structure $|\psi_{gs}\rangle$ not on the site basis as we did in Fig.~(\ref{sup}), but instead onto alternative basis states:  a basis $|\phi_m\rangle$ made of the eigenstates of a group of 13 coupled spirals. The idea is to take a microtubule segment which we can divide in multiples of 13 spirals and analyze which eigenstates of a block of 13 spirals contribute to form the superradiant ground state of the whole microtubule. Note that 13 is the minimum number of spirals we need to have a superradiant ground state (see Fig.~(\ref{f4}d)) and each block is made of $n_B=104 \times 13= 1352$ states.
If we call $|\psi_q^s \rangle$ the eigenstate $q$ of the $s$ block of 13 spirals, then the basis state $|\psi_m \rangle = |\psi_q^s \rangle$ for $(s-1) n_B < m \le s n_B$ while $|\psi_m \rangle = 0$ for $m > s n_B$ or $m \le (s-1) n_B$. In the upper panel of Fig.~(\ref{Proj}), the first $13 \times 104=1352$ states correspond to the eigenstates of the first 13 coupled spirals, the second $13 \times 104$ states correspond to the eigenstates of the coupled spirals from 14 to 26, and so on.
As one can see from Fig.~(\ref{Proj}), the components of the ground state over the 13 coupled spirals eigenstates are mainly concentrated in the ground states of each block of 13 coupled spirals (see also inset of Fig.~(\ref{Proj}) upper panel). 
The result in Fig.~(\ref{Proj})  clearly
shows that the ground state of the whole structure mainly consists
of a superposition of ground states of smaller blocks of spirals. 
This non-trivial result arises from the symmetry of the systems, see discussion in Ref.~\cite{gulli}. A very interesting consequence of this is that the total ground state emerges from coupling between the ground states of smaller blocks. Such coupling is of a supertransfer kind as we show below.

The supertransfer coupling~\cite{srlloyd} between the ground states of smaller blocks originates from the interaction of the giant dipole moments associated with the superradiant ground states of each block of  13 spirals. Indeed, the ground state of a block of 13 coupled spirals has a decay rate which is $\sim 235$ times larger than the single molecule decay rate.
In order to prove the previous statement,  let us compute the coupling  between the eigenstates  of two blocks of 13 coupled spirals, say block 1 and block 2. We will compute the coupling as a function of the distance between the two blocks, assuming the blocks are translated along the principal cylinder axis. 
Let us indicate the two corresponding $q$th eigenstates of the two blocks as $$|\psi^{s,q} \rangle= \sum_k C_k^{s,q} |k\rangle,$$ where the states $|k \rangle$ represent  the site basis of a block  and $s=1,2$.  The coupling between two single block eigenstates can be written as
\begin{equation}
\label{eq:st1}
V^q_{12} = \langle \psi^{1,q}| V| \psi^{2,q } \rangle= \sum_{k,k'} (C^{1,q}_k)^* C^{2,q}_{k'} V_{k,k'}.
\end{equation}
Note that $\langle \psi^{1,q}|$ is not the complex conjugate of $| \psi^{2,q } \rangle$ but the transpose of it, as we explain in Section 2. 
Using  Eqs. (\ref{Hmuk}),(\ref{eq:d1}), and (\ref{eq:g1}), we have  that  $V_{k,k'}=\Delta_{k,k'} -i\Gamma_{k,k'}/2= f(r_{k,k'}) \vec{\mu}_k \cdot \vec{\mu}_{k'} + g(r_{k,k'}) (\vec{\mu}_k \cdot \hat{r}_{k,k'}) (\vec{\mu}_{k'} \cdot \hat{r}_{k,k'})$, where the functions $f,g$ can be derived from Eqs. (\ref{Hmuk}),(\ref{eq:d1}), and (\ref{eq:g1}). When the distance between two blocks is much larger than their diameter we can approximate $r_{k,k'} \approx R_{12}$ where $R_{12}$ is the distance between the centers of the two blocks.  Eq.~(\ref{eq:st1}) then  becomes
\begin{equation}
\label{eq:st}
V^q_{12} =  \sum_{k,k'} (C^{1,q}_k)^* C^{2,q}_{k'} \left[ f(R_{12})\vec{\mu}_k \cdot\vec{\mu}_{k'} + g(R_{12})   (\vec{\mu}_k\cdot \hat{R}_{12}) (\vec{\mu}_{k'}\cdot \hat{R}_{12})\right],
\end{equation}
where $\vec{\mu}_{k}$  is the dipole moment of the $k$ molecule.
The above expression can be re-written in terms of the dipole strength of the eigenstates. 
The transition  dipole moment $\vec{D}_q$ associated with the $q$th eigenstate can be defined as follows: 
\begin{equation} \label{eq:dipst} 
\vec{D}_q=\sum_{i=1}^{N} C_{q,i} \, \vec{\mu}_i. 
\end{equation} 
The dipole coupling strength (often referred to as simply the dipole strength) of the $q$th eigenstate is defined  by $|\vec{D}_q|^{2}$ (note that due to normalization $\sum_{n=1}^{N}\left |\vec{D}_q\right|^{2}=N$).  Under the approximation that the imaginary part of the Hamiltonian (\ref{Hmuk}) can be treated as a perturbation and $L/\lambda \ll 1$ we have $|\vec{D}_q|^2 \approx \Gamma_q/\gamma$ (see \ref{app-b}). 
Thus, using Eq.~(\ref{eq:dipst}), Eq.~(\ref{eq:st})  can be re-written as
\begin{equation}
\label{eq:st2}
V^q_{12} =  \left[ f(R_{12}) |\vec{D_q}|^2  + g(R_{12})   (\vec{D}_q\cdot \hat{R}_{12}) (\vec{D}^*_{q}\cdot \hat{R}_{12})\right].
\end{equation}

 As a result for the coupling between the ground states of blocks of 13 spirals, we obtain $ V^{gs}_{12} \propto |D_{gs}|^2 \approx \Gamma_{gs}/\gamma \approx 235$, $\Gamma_{gs}/\gamma$ is the decay with of the ground state of 13 spirals (note that we can use the $|D_{gs}|^2 \approx \Gamma_{gs}/\gamma$  approximation  since for a block of 13 spirals we have $L/\lambda \approx 0.4$).
The above expression  represents the interaction between the giant dipoles of the ground states of each block. Therefore, states with a large dipole strength will have a supertransfer coupling proportional to the dipole strength of the eigenstates. Note that the coupling between eigenstates with a small dipole strength can give rise to a subtransfer effect, which has been shown in Ref.~\cite{gulli}. In the lower panel of Fig.~(\ref{Proj}), the coupling between the ground states  with a large dipole strength (blue circles) and between the most excited states with a very small dipole strength (green triangles) of two blocks of 13 spirals is compared with the average coupling between the molecules of each block (red squares). Note that the ground state of a block of 13 spirals is the most superradiant state with $\Gamma/\gamma \approx 235$, while the highest-energy state is the most subradiant with the lowest decay width $\Gamma/\gamma \approx 10^{-6}$ for a block of 13 spirals. The couplings are shown as a function of the center-to-center distance between the two blocks normalized by the wavelength connected with the optical transition. One can see that the coupling between the ground states is significantly larger than the average coupling between the molecules. Moreover, for large center-to-center distances $d$, the coupling between the ground states is well-approximated by Eq.~(\ref{eq:st2}) (see blue dashed curve), thus proving the existence of a supertransfer effect. On the other hand, the coupling between the most excited states of the two blocks is much smaller than the average coupling between the molecules, showing a subtransfer effect. 
The above results suggest that the dynamics will be very dependent on the initial conditions and will exhibit at least two distinct timescales due to the presence of supertransfer and subtransfer processes.
In the next section, we will show how the cooperativity-enhanced coupling between the ground states of blocks of spirals can boost photoexcitation transport.

\section{Transport of photoexcitations via supertransfer}
In this section we consider the spreading velocity of an initial excitation concentrated
in the middle of a microtubule made of 99 spirals, with a total length of about $800$ nm.
The spreading of the initial excitation has been measured by the root-mean-square deviation (RMSD) of the excitation along the longitudinal axis of the microtubule. Given that the initial state of the system is described by the wavefunction $|\psi(0) \rangle$,
the average position of the excitation on the axis of the microtubule can be computed  with $\overline{Q}(t)= \sum_k |\langle k|\psi(t) \rangle|^2 z_k$, where $z_k$ is the position of the $k$-th molecule on the longitudinal axis and $ |\psi(t) \rangle$ is the wavefunction at time $t$.  Thus the variance as a function of time can be computed with $\sigma^2(t)= \sum_k |\langle k|\psi(t) \rangle|^2 z_k^2 - \overline{Q}(t)^2$, from which follows $\text{RMSD}(t)=\sqrt{\sigma^2(t)}$.

\begin{figure}
    \centering
\includegraphics[ trim=1cm 0 12 0,scale=0.45]{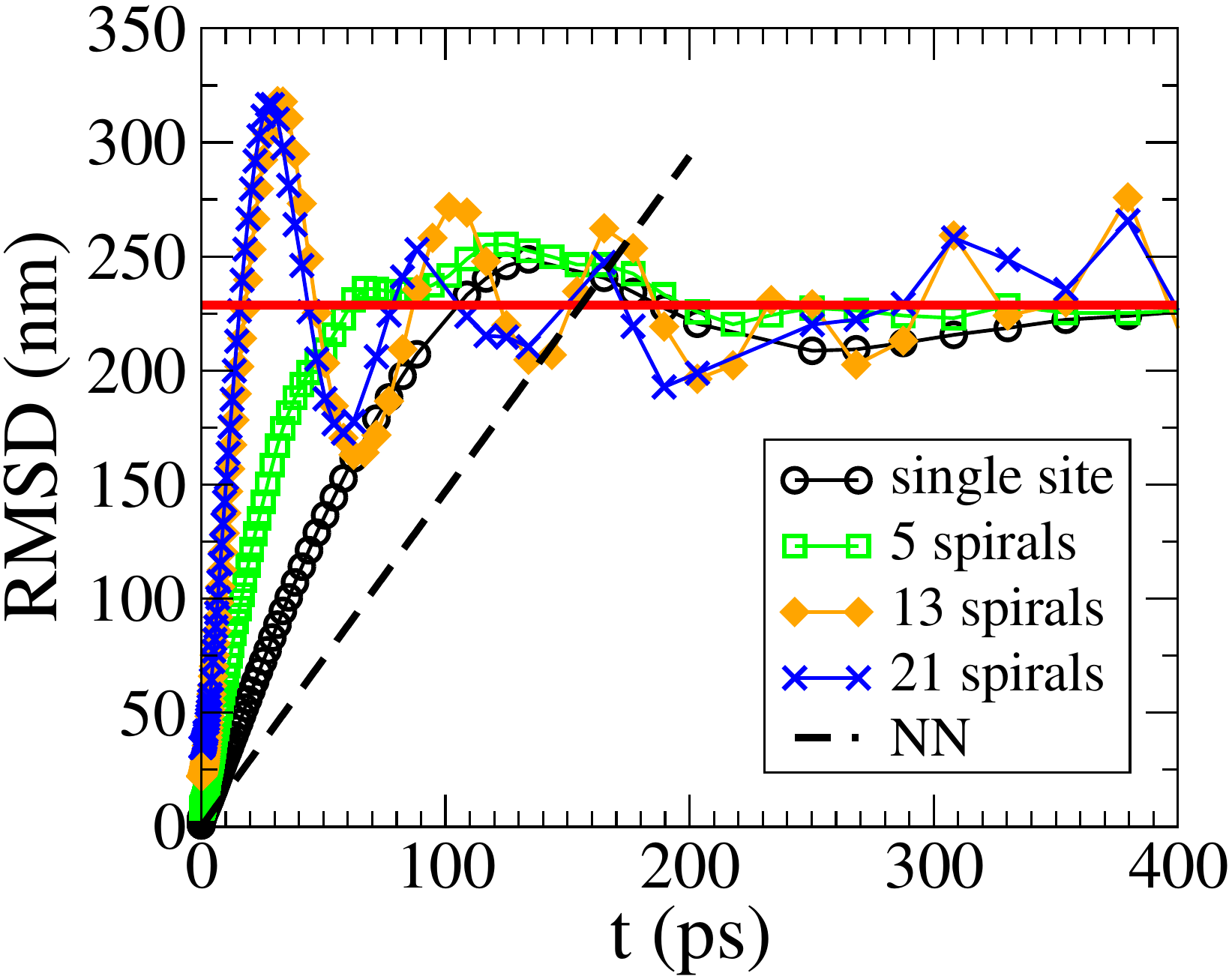} \hspace{0.5cm}
\includegraphics[ trim=1cm 0 12 0,scale=0.45]{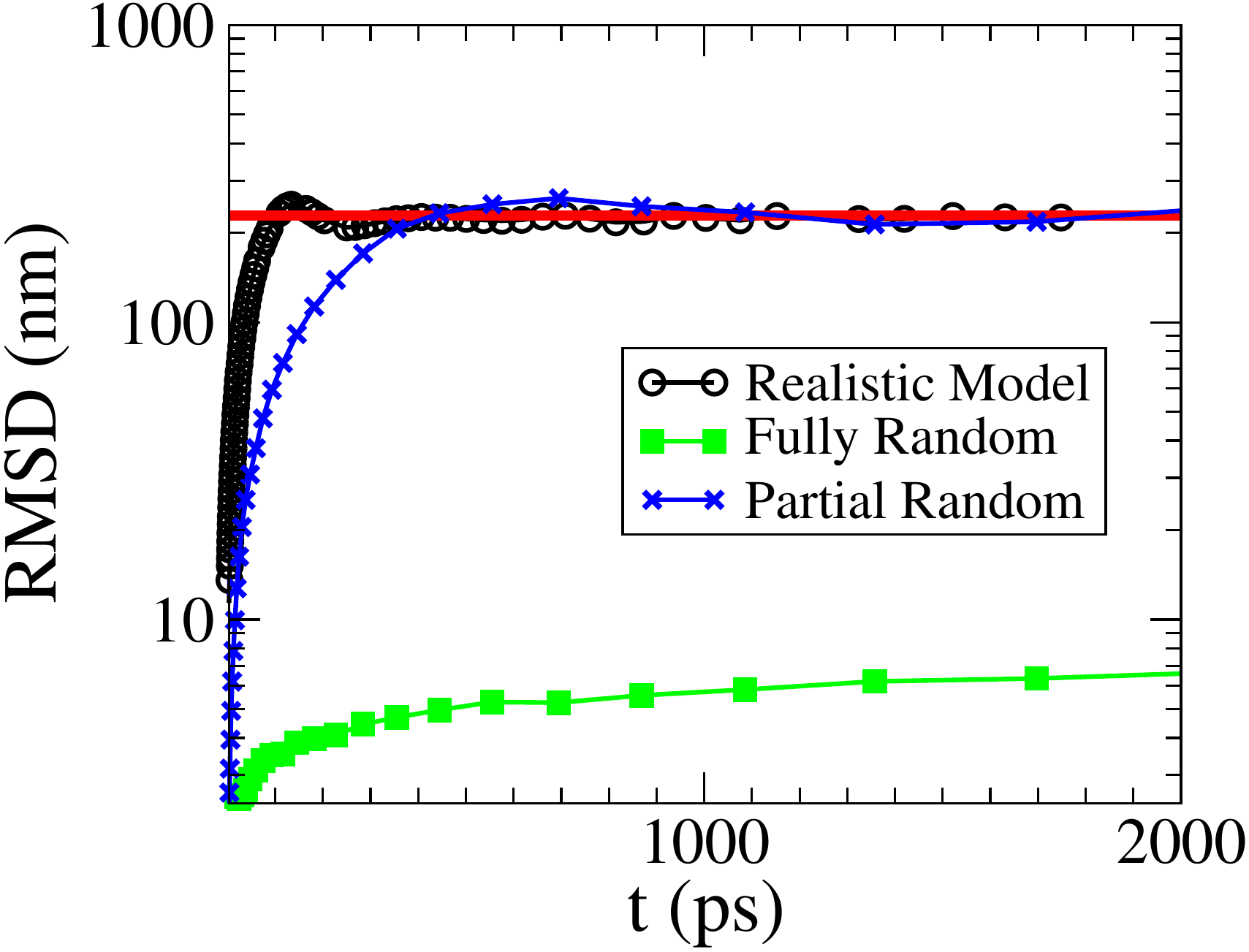}
\caption{ \\Root-mean-square deviation (RMSD) as a function of time for an initial excitation concentrated in the middle of a microtubule comprised of 99 spirals with a total length of about $810$ nm. Left panel: The case of an initial excitation concentrated on a single site of the central spiral (black circles) is compared with an excitation located on the ground state of the central five (empty green squares), 13 (orange crosses), and 21 (blue stars) spirals. For comparison, the spreading expected from the strength of the  nearest-neighbor coupling is also shown (black dashed line) and described in the text. The equilibrium value for an excitation equally distributed over all the tryptophans of the microtubule is shown as a red horizontal line. The spreading velocity of an excitation starting from a single site is about two times faster than the spreading associated with the amplitude of the nearest-neighbor coupling (NN velocity), while the spreading velocity of an excitation which starts in the ground state of more than 13 spirals is about ten times faster than the NN velocity. Right panel: Here the case of an initial excitation concentrated on a single site of the central spiral is considered for three different models - the realistic model (black circles and same data as in the left panel), the fully random model where all the dipoles are randomly oriented (filled green squares), and the partial random model where the dipoles are oriented at random but repeated in the same configuration for all spirals (blue crosses).}
\label{RMS}
\end{figure}

We have chosen different initial conditions to show the effect of cooperativity on the spreading
of the excitation: $(i)$ an initial excitation concentrated on a single randomly selected site of the central spiral; 
and $(ii)$ an initial excitation concentrated in the ground state of the central block of 5, 13, or 21 spirals. As displayed in Fig.~(\ref{RMS}),
the spreading of the initial wave packet is always ballistic ($\text{RMSD}(t) \propto t$)
so that we can define a velocity of spreading $V$ as the linear slope.
Note that $V$ increases as we increase the number of spirals over which the initial excitation
is spread and then saturates when the number of spirals becomes large. Indeed
from  Fig.~(\ref{RMS}) we can see that the spreading velocity is the same when the initial state coincides
with the ground state of 13 or 21 central spirals.
When the excitation starts from the ground state of 21 central spirals, $V$ is more
than five times the velocity of an excitation concentrated on a single site.
Such an effect is due to supertransfer coupling between the ground states of blocks of spirals, and as a consequence of the fact that the ground states of the central spirals
have a large overlap with the extended superradiant ground state, as shown in  Fig.~(\ref{Proj}).

For comparison we also estimated  the spreading velocity of an excitation which can be expected based on the typical nearest-neighbor coupling present in the system.
In the Trp case the typical nearest-neighbor coupling is $J \approx 50 \text{ cm}^{-1}$, so that
the time needed for the excitation to move by one Trp can be estimated as 
$\tau= 1/ (4 \pi c J)  \approx 0.053$ ps, where the light velocity is $c \approx 0.03$ cm/ps. This can be derived from the period of oscillation between two sites at resonance, which is given by $T=\hbar/J'= hc/(2 \pi c J')$, and $\tau=T/2$ (note that $J=J'/(hc)$ is the coupling in cm$^{-1}$ as measured in this paper). The average distance between Trps projected along the main axis of the microtubule can be evaluated from $d=810 \text{ nm}/10400=0.078$ nm. We thus obtain a velocity $V_{NN}=d/\tau \approx 1.47$ nm/ps, which is  shown by the black dashed line in
Fig.~(\ref{RMS}). Note that $V_{NN}$ is about two times smaller than the spreading velocity starting from a single site. This is probably due to the fact that the long-range interactions between
the sites favor the spreading of the excitation (see Section 5). Most importantly, $V_{NN}$ is 
 about ten times smaller then the spreading velocity of a delocalized excitation obtained by setting the initial state equal to the ground state of 13 or more central spirals. 

For large times, the RMSD reaches a stationary value that assumes the excitation becomes equally distributed on all Trps
of the microtubule. We can compute such a stationary value of the RMSD
from the positions of the Trps, by setting $\overline{Q}= \sum_k z_k/N$
and $\sigma^2= \sum_k  z_k^2/N - \overline{Q}^2$, so that  $\text{RMSD}=\sqrt{\sigma^2}$.
For a microtubule made of 99 spirals we obtain RMSD $\approx 228$ nm.
This value is shown in Fig.~(\ref{RMS}) as a horizontal red line, and
it agrees very well with the numerical results.

We would like to emphasize that a photon is likely to be absorbed by the ground state
of a block of spirals, since the ground state is the state which is most strongly coupled to
the electromagnetic field (i.e., it has the highest decay width and absorption rate). For this reason an initial excitation coinciding with the
ground state of a block of spirals is well motivated, and the fact that its spreading is enhanced can have important consequences for photoexcitation transport.

Finally, in order to emphasize the role of symmetry in the transport properties of the system, in the right panel of Fig.~(\ref{RMS})  we show the spreading of an initial excitation starting from a single site on the central spiral for the realistic model considered before (black circles in both left and right panels represent the same data), for the fully random model, and for the partial random model. In both the latter models, the positions of the dipoles are kept fixed with respect to the realistic model, but the orientation of their dipoles has been randomized. Note that in the fully random model the dipole directions have been randomized along the entire microtubule length, whereas for the partial random model we have randomized the dipole orientations in one spiral and then repeated this configuration in all other spirals. 
For the partial random model, the excitation spreads over the whole microtubule segment with a smaller velocity than the realistic model, while for the fully random model the spreading is extremely slow and remains well below the value (see horizontal red line) of an equally distributed excitation during the whole simulation time. The above results show the relevance of native symmetry in excitonic energy transport through the microtubule.

\section{Robustness to disorder and the role of long-range interactions} 

In order to study the robustness to disorder of the superradiant ground
state, we have analyzed the system in the presence of static disorder, i.e., time-independent
and space-dependent fluctuations of the excitation energies of the tryptophans
comprising the microtubule chromophoric lattice. Specifically we consider that the excitation energies of the tryptophans
are uniformly distributed around the initial value $e_0$, between $e_0-W/2$ and $e_0+W/2$,
so that $W$ represents the strength of the static disorder. 
It is well known that static disorder induces localization of the eigenstates of a
system, a phenomenon known as Anderson localization~\cite{Anderson}. Due to such localization,
for each eigenstate the probability of finding the excitation is concentrated
on very few sites for large disorder, and only on one site for extremely large disorder. Note that Anderson localization usually occurs in the presence of short-range interactions, but in our model there are multiple contributions from a complicated power law for the interaction (see Eq. (\ref{eq:d1})), so that the results of our analysis are not obvious.

\begin{figure}[ht!]
    \centering
\includegraphics[ trim=1cm 0 12 0,scale=0.6]{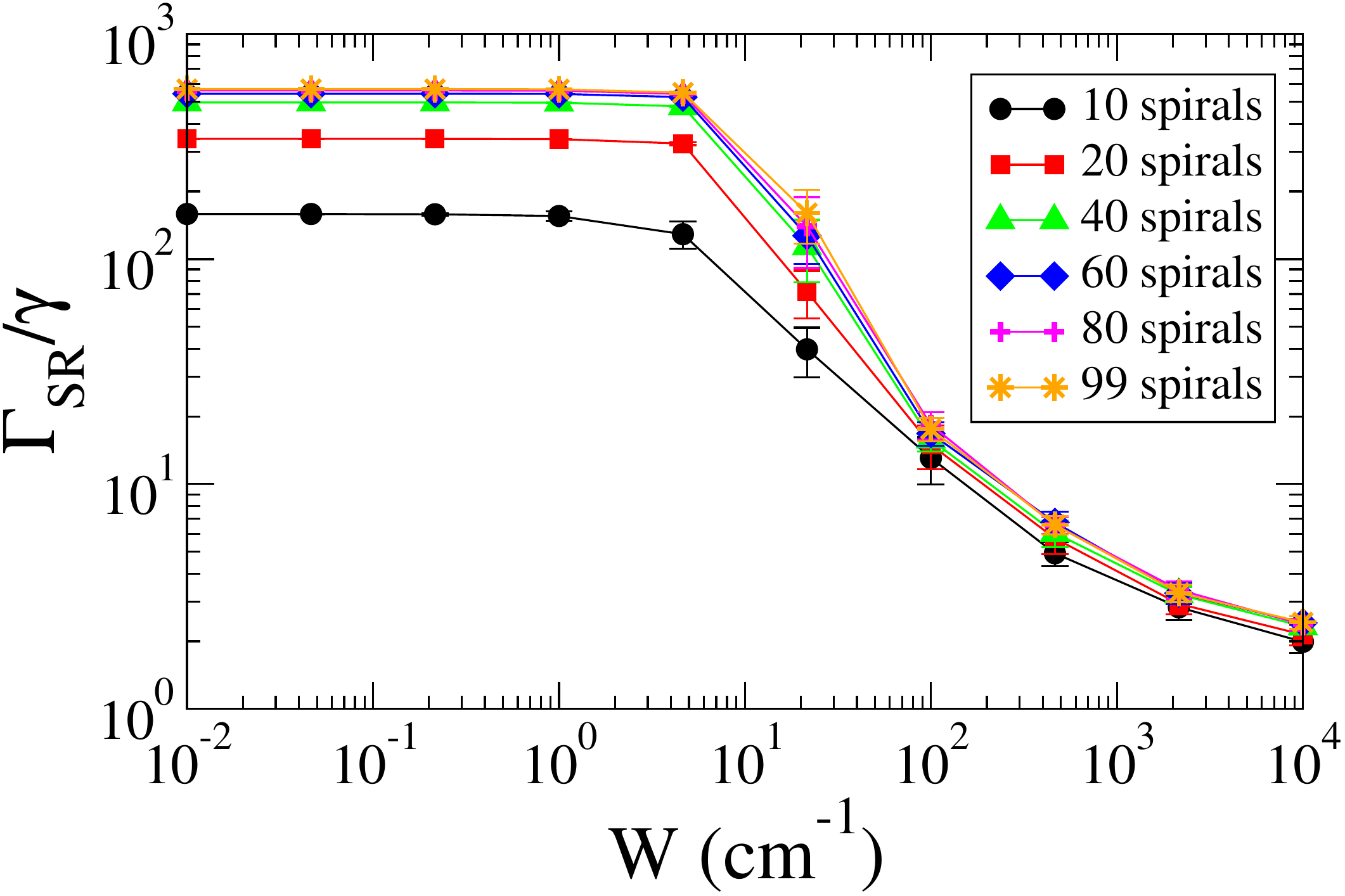}
\caption{ Normalized decay width of the superradiant state is plotted vs. the strength
  of static disorder $W$ for a microtubule segment of different lengths.}
\label{GW}
\end{figure}

In order to study the robustness of superradiance to disorder, we plot in 
Fig.~(\ref{GW}) the maximum normalized decay width $\Gamma_{SR}/\gamma$  as a function of the disorder strength $W$, using the full realistic non-Hermitian Hamiltonian given in Eq.~(\ref{Hmuk}) for different microtubule sizes. 
Note that the most superradiant state (i.e., largest decay width) coincides with the ground state of the microtubule for sizes larger than 12 spirals. 
One can see that the disorder at which the width
of the superradiant state starts to decrease is independent of the system size (within the system sizes considered in our simulations). 
This is quite surprising for quasi-one dimensional structures, which usually exhibit a critical disorder that decreases as the system size grows~\cite{beenakker}. 
Indeed, for short-range interactions in quasi-one dimensional structures, the critical  static disorder strength $W_{cr}$ needed to localize the system  goes to zero as the system size goes to infinity, with $W_{cr} \propto J/\sqrt{N} $.


\begin{figure}[ht!]
    \centering
\includegraphics[ trim=1cm 0 12 0,scale=0.6]{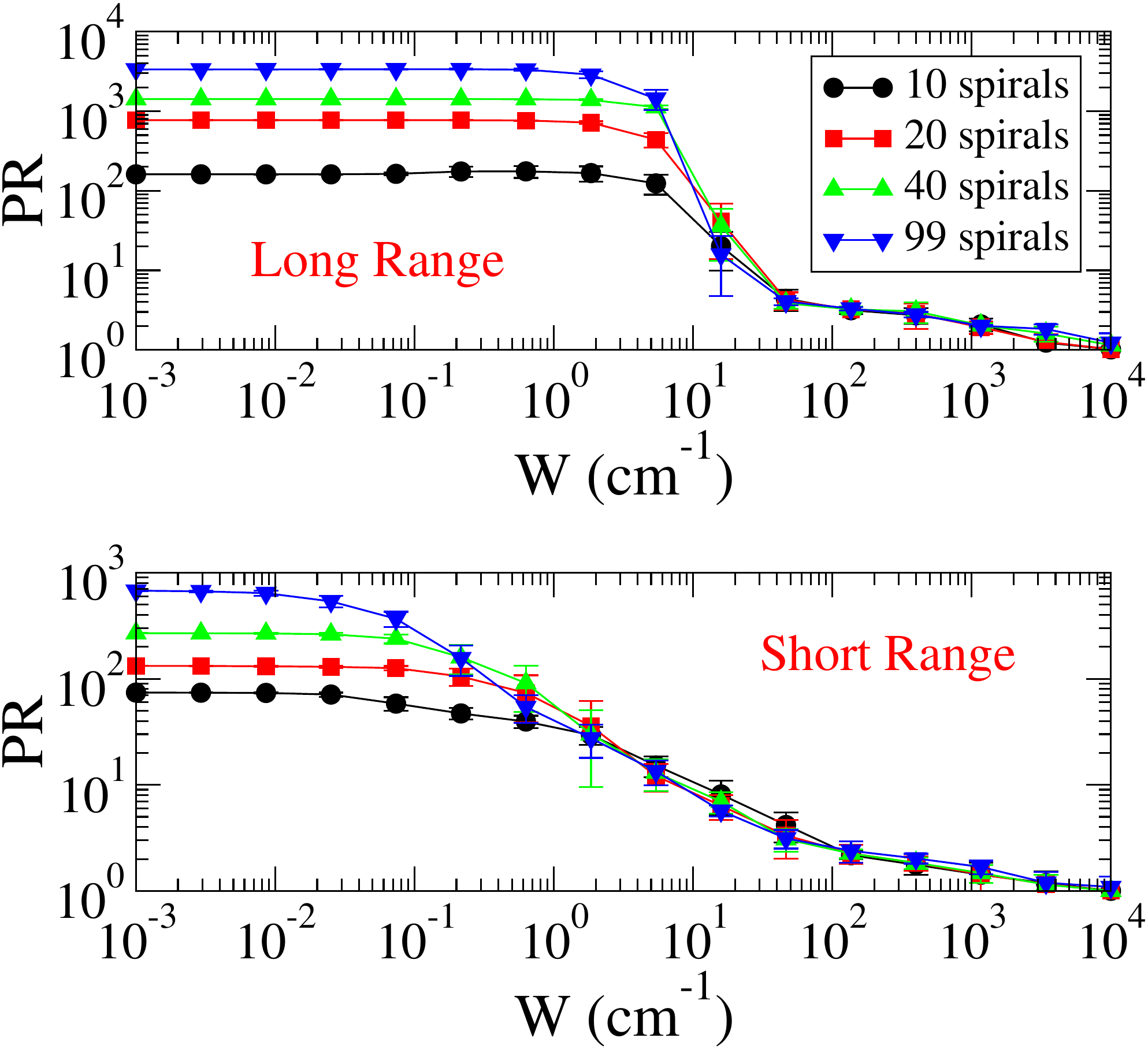}
\caption{Participation ratio (PR) of the ground state for a microtubule of different lengths is shown vs. the strength of static disorder $W$, averaged over ten realizations of disorder for each system length. The upper panel presents the results of the full model, where the interaction between the tryptophans has a long-range nature (see Eq.~(\ref{Hmuk})). However, note that only the Hermitian part of the Hamiltonian in Eq.~(\ref{Hmuk}) has been used to obtain the data shown in both panels of this figure. In the lower panel, the PR of the ground state for a microtubule of different lengths is shown vs. the strength of static disorder $W$ for the case of short-range interactions only, considering only the couplings between tryptophans with a center-to-center separation smaller then $2$ nm. 
} 
\label{PRW}
\end{figure} 

In order to understand how the above results could be explained by the effective range of the interaction, we have compared our realistic model which contains long-range interactions with the same model where the long-range interactions have been suppressed. Specifically the short-range model has been obtained by considering only the interactions between Trps with a center-to-center separation less than $2$ nm. In order to perform such a comparison, only the Hermitian part of the realistic model Hamiltonian given in
Eq.~(\ref{Hmuk}) has been taken into account. We considered only the Hermitian part of the Hamiltonian because it 
constitutes a good approximation of the whole Hamiltonian (see discussion in \ref{app-b}) and, most importantly, allows for comparison of different ranges of interaction. Indeed, in the full non-Hermitian model we cannot change the range of the interaction without introducing inconsistencies (i.e., negative decay widths). 
Below we will show that the results thus obtained are consistent with the analysis of the whole Hamiltonian given in Fig.~(\ref{GW}), where the full non-Hermitian Hamiltonian has been considered.

We analyzed the effect of disorder for the two different models (long range and short range)  through the participation ratio (PR) 
\begin{equation}
 \text{PR}= \left\langle              
\sum_i |\langle q_i| \psi \rangle|^2 \bigg/ 
\sum_i |\langle q_i| \psi \rangle|^4  
\right\rangle
 \label{pr}
\end{equation}
of the eigenstates $|\psi \rangle$  of the system, 
where  the large outer brackets, $\langle {\dots} \rangle$, stand for the ensemble average over 
different realizations of the static disorder.
The PR is widely used to characterize localization properties~\cite{pr},
and it satisfies the bounds $1\le \text{PR}\le N$.
For extended states, it increases proportionally to the system size $N$,
while, for localized states, it is independent of $N$. 

In order to study the effect of static disorder,
we have analyzed the PR of the ground state as a function of the disorder strength $W$ in
Fig.~(\ref{PRW}).  As shown in the upper panel of Fig.~(\ref{PRW}), where the long-range model is analyzed, the critical disorder at which the PR starts
to decrease appears to be independent of the microtubule length. 
The critical disorder obtained in this case is consistent with the analysis of the critical disorder needed to quench superradiance as shown in Fig.~(\ref{GW}). 
The response of the system to disorder is completely different for the short-range model. The robustness to disorder of the ground state of such a model is shown in the lower panel of Fig.~(\ref{PRW}). As one can see, the critical disorder decreases with the system size in this case, as would be expected for quasi-one dimensional systems with short-range interactions.
The difference between the two panels of Fig.~(\ref{PRW}) shows that the long-range
nature and symmetry of the interactions play a very significant role in 
enhancing the robustness of excitonic states in microtubules to disorder.
However, we note that robustness to disorder could also be connected to supertransfer, and not only to the long range of the interaction. For further details see the discussion in Ref.~\cite{gulli}  and in \ref{app-gap}.

\section{Conclusions and perspectives}
We have analyzed the excitonic response of microtubules induced by the coupling of
tryptophan molecules, which are the most strongly photoactive molecules in the spiral-cylindrical lattice. The positions and
orientations of the dipoles of the tryptophan molecules have been obtained in
previous works by molecular dynamics simulations and quantum chemistry calculations and have closely reproduced experimental spectra for the tubulin heterodimeric protein ~\cite{kurian}. Analyzing the properties of 
a microtubule of length $L \sim 800$ nm, which is larger than the wavelength of the excitation transition ($L > \lambda = 280$ nm), requires an approach that goes beyond the transition dipole-dipole couplings alone. This is why we take into consideration radiative interactions containing non-Hermitian terms.

Our analysis has shown that
the coupling between tryotophan molecules is able to create a 
superradiant ground state, similar to the physical behavior of several photosynthetic antenna systems. 
Such a superradiant ground state, which absorbs in the UV spectral range, has been shown to be a
coherent excitonic state extended over the whole microtubule lattice of tryptophan molecules. Interestingly, the superradiant ground state appears to be delocalized on the exterior wall of the microtubule, which interfaces with the cytoplasm, suggesting the possibility that these extended but short-lived (few picosecond) excitonic states may be involved in communication with cellular proteins that bind to microtubules in order to carry out their functions. At the same time, we have shown that long-lived (hundreds of milliseconds) subradiant states can be concentrated on the inner wall of the microtubule lumen, potentially maintaining excitonic transfer processes across the cytoskeletal network in a more ``protected'' thermodynamic milieu. These subradiant states could be particularly important in the synchronization of neuronal processes in the brain, where microtubules can extend to the micron scale and beyond. 

Our analysis may have further biological implications. In a series of studies spanning a period of almost a quarter century, G. Albrecht-Buehler observed that living cells possess a spatial orientation mechanism located in the centrosome \cite{Albrecht-Buehler,AB1,AB2,AB4}. The centrosome is formed from an intricate arrangement of microtubules in two perpendicular sets of nine triplets (called centrioles). This arrangement provides the cell with a primitive ``eye'' that allows it to locate the position of other cells within a two-to-three-degree accuracy in the azimuthal plane and with respect to the axis perpendicular to it \cite{AB4}. While Albrecht-Buehler proposed that centrosomes are infrared detectors, it is still a mystery how the reception of electromagnetic radiation is accomplished by the centrosome. Superradiant behavior in these microtubule aggregates may play a role.

Moreover, we have shown that the superradiant ground state of the whole microtubule arises through a supertransfer coupling between the ground states of smaller blocks of spirals within the microtubule. For this reason,  microtubule superradiance is essentially an emergent property of the whole system that develops as ``giant dipole'' strengths of superradiant ground states of constituent blocks interact to form a delocalized coherent state on the entire structure. This is a hallmark of self-similar behavior, in the sense that subunit blocks of spirals exhibit superradiant characteristics that recapitulate roughly what is seen in the whole. Only by considering the entire structure (or at least a substantial fraction of the spirals) do we uncover cooperative and dynamical features of the system that would otherwise fail to be captured in more reductionist models. Supertransfer coupling between excitonic states of different blocks of spirals in the microtubule segment is critical to the manifestation of these cooperative behaviors and explains the calculated couplings to excellent agreement.

Such supertransfer coupling is able to enhance the spreading of photoexcitation inside the microtubule. The spreading of photoexcitation
is ballistic, despite the fact that the native dipole orientations of the tryptophan molecules are not fully symmetric even in the absence of static disorder (see Fig. (\ref{figMT}b)). The spreading velocity is strongly
dependent on the initial conditions, and, due to supertransfer, it can be about ten times faster
than the velocity expected from the amplitude of the nearest-neighbor coupling
between the tryptophan molecules in such structures. These results show that the characteristic supertransfer processes analyzed in photosynthetic antenna complexes may also be present in microtubules.

Finally, we have analyzed the robustness of microtubule superradiance to static disorder. 
We have shown that the symmetry and long-range nature of the interactions
give an enhanced robustness to such structures with a critical disorder which appears to be
independent of the system size (up to the system sizes analyzed in this paper). This is at variance with  what usually happens in quasi-one dimensional
structures with short-range interactions, where the critical disorder $W_{cr}$ goes to zero
as the system size grows. Indeed, for quasi-one dimensional systems with short-range interactions only, we have $W_{\rm cr} \propto J/\sqrt{N}$, where $J$ is the  nearest-neighbor coupling and $N$ is the number of chromophore sites. 
The critical disorder at which superradiance and the delocalization
of the ground state are precipitously affected is found to be on the order of $10 \text{ cm}^{-1}$ (see Fig.~(\ref{PRW}a)). Such a value of disorder is not extremely large, as natural disorder can be on the order of $kT$, ranging from $50 \text{ cm}^{-1}$ to $300 \text{ cm}^{-1}$. 
Still, such critical disorder is much larger than the critical disorder expected from the typical nearest-neighbor coupling between tryptophan molecules. Indeed, for a microtubule of
$\sim 800$ nm with only nearest-neighbor interactions containing $N \simeq 10^4$ molecules, one would obtain a critical disorder of about $10^{-1} \text{ cm}^{-1}$ (corresponding to 50 cm$^{-1}/\sqrt{N}$), two orders of magnitude smaller than what we have found. 
Such enhanced robustness to static disorder as a result of long-range interactions and symmetry can greatly increase diffusion lengths and thereby support ultra-efficient photoexcitation transport.

To refine our studies, future work should certainly include consideration of the other photoactive amino acids present in microtubules and the effects of thermal relaxation on coherent energy transport.
The significance of photoexcitation transport in microtubules is an
open question in the biophysics community, and further
experimental and theoretical works are needed to establish the precise mechanisms of their optical functionality. Our results point towards a possible role of superradiant  and supertransfer processes in microtubules. Both cooperative effects are able to induce ultra-efficient photoexcitation absorption and could serve to enhance energy transport over long distances under natural conditions. We hope that our results will inspire further experimental studies on microtubules to gather evidence for UV superradiance in such important biological structures.

\ack  
GLC acknowledges the support of PRODEP (511-6/17-8017). PK was supported in part by the National Institute on Minority Health and Health Disparities of the National Institutes of Health under Award Number G12MD007597. The content is solely the responsibility of the authors and does not necessarily represent the official views of the National Institutes of Health. PK would also like to acknowledge support from the US-Italy Fulbright Commission and the Whole Genome Science Foundation. TJAC would like to acknowledge financial support from the Department of Psychology and Neuroscience and the Institute for Neuro-Immune Medicine at Nova Southeastern University (NSU), and work in conjunction with the NSU President's Faculty Research and Development Grant (PFRDG) program under PFRDG 335426 (Craddock – PI).

\appendix
\section{Microtubule tryptophan dipole positions and orientations}
\label{app-a}
The tubulin $\alpha$-$\beta$ heterodimer structure was obtained by repairing the protein data bank (PDB: www.rcsb.org) \cite{PDB} structure 1JFF \cite{1JFF} by adding missing residues from 1TUB \cite{1TUB} after aligning 1TUB to 1JFF. This initial repaired dimer structure was oriented by itself alone such that the (would-be) protofilament direction aligned with the $x$-axis, the normal to the (would-be) outer microtubule surface aligned with the $y$-axis, and the direction of (would-be) lateral contacts aligned with the $z$-axis, before subsequent translation and rotation.  A single spiral of 13 tubulin dimers was generated by translating each dimer 11.2 nanometers (nm) in the $y$-direction, then successively rotating the resulting dimer structure by multiples of -27.69$^\circ$ in the $y$-$z$ plane about the origin around the $x$-axis, and successively shifting each dimer by multiples of 0.9 nanometers in the $x$-direction.  This resulted in a left-handed helical-spiral structure with a circular radius of 22.4 nm passing through the center of each dimer in the B-lattice microtubule geometry described by Li et al. \cite{Li2002} and Sept et al. \cite{Sept2003}. The orientation of the $1L_a$ excited state of each tryptophan molecule in the resulting structure was taken as 46.2$^\circ$ above the axis joining the midpoint between the CD2 and CE2 carbons of tryptophan and carbon CD1, in the plane of the indole ring (i.e., towards nitrogen NE1). The Cartesian positions and unit vector directions of the 104 tryptophans of the first spiral are given in Table A1 below. To generate successive spirals, the initial spiral coordinates were translated along the $x$ (i.e., protofilament) direction by multiples of 8 nm. Modeling was done with PyMOL 1.8.6.2 \cite{PYMOL}. 

\appendix
\begin{table}
\setcounter{section}{1}
\caption{First Spiral Dipole Positions (\AA) and Unit Vectors.}
\lineup
\begin{indented}
\item[]\begin{tabular}{@{}llllll}
\br
$x$	&	$y$	&	$z$	&	$\hat{\mu}_x$	&	$\hat{\mu}_y$	&	$\hat{\mu}_z$	\\
\mr
-2.378	&	103.218	&	14.720	&	-0.70114	&	0.66510	&	-0.25699	\\
-13.691	&	123.899	&	7.109	&	0.65456	&	-0.70298	&	-0.27816	\\
13.384	&	124.916	&	-9.487	&	-0.53855	&	-0.00510	&	-0.84258	\\
-28.566	&	122.415	&	5.686	&	-0.18573	&	-0.21751	&	-0.95822	\\
6.622	&	97.563	&	-31.783	&	-0.70125	&	0.46922	&	-0.53674	\\
-4.691	&	112.339	&	-48.133	&	0.65455	&	-0.75173	&	0.08039	\\
22.384	&	105.527	&	-63.300	&	-0.53854	&	-0.39597	&	-0.74376	\\
-19.566	&	110.363	&	-48.703	&	-0.18568	&	-0.63815	&	-0.74718	\\
15.622	&	70.946	&	-70.331	&	-0.70134	&	0.16574	&	-0.69329	\\
4.309	&	76.431	&	-91.675	&	0.65456	&	-0.62818	&	0.42064	\\
31.384	&	63.350	&	-101.939	&	-0.53842	&	-0.69643	&	-0.47443	\\
-10.566	&	74.417	&	-91.261	&	-0.18563	&	-0.91228	&	-0.36509	\\
24.622	&	29.463	&	-92.094	&	-0.70142	&	-0.17512	&	-0.69090	\\
13.309	&	24.401	&	-113.542	&	0.65455	&	-0.36084	&	0.66435	\\
40.384	&	8.049	&	-116.552	&	-0.53845	&	-0.83715	&	-0.09615	\\
-1.566	&	22.810	&	-112.239	&	-0.18584	&	-0.97746	&	0.10022	\\
33.622	&	-17.382	&	-92.086	&	-0.70121	&	-0.47630	&	-0.53052	\\
22.309	&	-31.831	&	-108.725	&	0.65455	&	-0.01098	&	0.75594	\\
49.384	&	-47.710	&	-103.791	&	-0.53840	&	-0.78605	&	0.30374	\\
7.435	&	-32.636	&	-106.832	&	-0.18561	&	-0.81878	&	0.54327	\\
42.622	&	-58.858	&	-70.309	&	-0.70112	&	-0.66844	&	-0.24825	\\
31.309	&	-79.384	&	-78.327	&	0.65455	&	0.34193	&	0.67427	\\
58.384	&	-91.151	&	-66.579	&	-0.53840	&	-0.55461	&	0.63445	\\
16.435	&	-79.217	&	-76.278	&	-0.18581	&	-0.47252	&	0.86151	\\
51.622	&	-85.462	&	-31.752	&	-0.70143	&	-0.70694	&	0.09067	\\
40.309	&	-107.363	&	-29.312	&	0.65456	&	0.61602	&	0.43825	\\
67.384	&	-112.323	&	-13.442	&	-0.53847	&	-0.19691	&	0.81932	\\
25.435	&	-106.263	&	-27.576	&	-0.18598	&	-0.01812	&	0.98239	\\
60.622	&	-91.101	&	14.753	&	-0.70119	&	-0.58426	&	0.40863	\\
49.309	&	-109.360	&	27.091	&	0.65454	&	0.74918	&	0.10153	\\
76.384	&	-106.376	&	43.448	&	-0.53846	&	0.20679	&	0.81688	\\
34.435	&	-107.578	&	28.118	&	-0.18579	&	0.44042	&	0.87836	\\
69.622	&	-74.482	&	58.551	&	-0.70122	&	-0.32715	&	0.63346	\\
58.309	&	-84.916	&	77.961	&	0.65457	&	0.71067	&	-0.25784	\\
85.384	&	-74.672	&	91.058	&	-0.53850	&	0.56268	&	0.62723	\\
43.435	&	-82.861	&	78.042	&	-0.18566	&	0.79824	&	0.57301	\\
78.622	&	-39.413	&	89.609	&	-0.70101	&	0.00458	&	0.71313	\\
67.309	&	-39.631	&	111.644	&	0.65460	&	0.50989	&	-0.55814	\\
94.384	&	-24.474	&	118.481	&	-0.53846	&	0.78961	&	0.29424	\\
52.435	&	-37.774	&	110.761	&	-0.18583	&	0.97305	&	0.13653	\\
87.622	&	6.073	&	100.812	&	-0.70134	&	0.33563	&	0.62887	\\
76.309	&	16.121	&	120.425	&	0.65458	&	0.19139	&	-0.73137	\\
103.384	&	32.718	&	119.434	&	-0.53849	&	0.83584	&	-0.10678	\\
61.435	&	17.355	&	118.780	&	-0.18556	&	0.92497	&	-0.33168	\\
96.622	&	51.555	&	89.594	&	-0.70137	&	0.58928	&	0.40104	\\
85.309	&	69.566	&	102.291	&	0.65456	&	-0.17071	&	-0.73649	\\
112.384	&	83.802	&	93.700	&	-0.53847	&	0.69044	&	-0.48305	\\
70.435	&	69.894	&	100.261	&	-0.18584	&	0.66506	&	-0.72330	\\
\br
\end{tabular}
\end{indented}
\end{table}

\appendix
\begin{table}
\setcounter{section}{1}
\caption{First Spiral Dipole Positions (\AA) and Unit Vectors. \textit{(cont.)}}
\lineup
\begin{indented}
\item[]\begin{tabular}{@{}llllll}
\br
$x$	&	$y$	&	$z$	&	$\hat{\mu}_x$	&	$\hat{\mu}_y$	&	$\hat{\mu}_z$	\\
\mr
105.622	&	86.614	&	58.524	&	-0.70131	&	0.70820	&	0.08138	\\
94.309	&	108.463	&	61.397	&	0.65456	&	-0.49312	&	-0.57304	\\
121.384	&	117.076	&	47.174	&	-0.53852	&	0.38715	&	-0.74841	\\
79.435	&	107.810	&	59.447	&	-0.18562	&	0.25246	&	-0.94964	\\
39.057	&	102.384	&	14.619	&	-0.01722	&	-0.52500	&	0.85093	\\
53.331	&	124.899	&	-8.386	&	-0.68314	&	0.49202	&	-0.53967	\\
36.453	&	121.233	&	-4.011	&	0.98886	&	0.14794	&	0.01663	\\
12.616	&	122.434	&	5.625	&	-0.19086	&	-0.22402	&	-0.95571	\\
48.057	&	96.778	&	-31.485	&	-0.01690	&	-0.06907	&	0.99747	\\
62.331	&	106.023	&	-62.318	&	-0.68315	&	0.18496	&	-0.70647	\\
45.453	&	104.810	&	-56.739	&	0.98882	&	0.13886	&	-0.05437	\\
21.616	&	110.352	&	-48.767	&	-0.19062	&	-0.64262	&	-0.74210	\\
57.057	&	70.389	&	-69.702	&	-0.01692	&	0.40217	&	0.91541	\\
71.331	&	64.247	&	-101.300	&	-0.68315	&	-0.16473	&	-0.71146	\\
54.453	&	65.765	&	-95.796	&	0.98882	&	0.09792	&	-0.11249	\\
30.616	&	74.376	&	-91.313	&	-0.19047	&	-0.91372	&	-0.35893	\\
66.057	&	29.262	&	-91.278	&	-0.01730	&	0.78143	&	0.62376	\\
80.331	&	9.139	&	-116.402	&	-0.68314	&	-0.47649	&	-0.55342	\\
63.453	&	13.041	&	-112.235	&	0.98885	&	0.03427	&	-0.14494	\\
39.616	&	22.750	&	-112.267	&	-0.19077	&	-0.97579	&	0.10698	\\
75.057	&	-17.181	&	-91.270	&	-0.01707	&	0.98186	&	0.18883	\\
89.331	&	-46.675	&	-104.165	&	-0.68312	&	-0.67918	&	-0.26845	\\
72.453	&	-41.283	&	-102.288	&	0.98885	&	-0.03702	&	-0.14422	\\
48.616	&	-32.701	&	-106.829	&	-0.19059	&	-0.81421	&	0.54839	\\
84.057	&	-58.300	&	-69.680	&	-0.01707	&	0.95720	&	-0.28891	\\
98.331	&	-90.408	&	-67.392	&	-0.68309	&	-0.72612	&	0.07834	\\
81.453	&	-84.762	&	-68.236	&	0.98889	&	-0.09973	&	-0.11025	\\
57.616	&	-79.273	&	-76.244	&	-0.19047	&	-0.46622	&	0.86392	\\
93.057	&	-84.676	&	-31.454	&	-0.01721	&	0.71339	&	-0.70055	\\
107.331	&	-112.043	&	-14.506	&	-0.68310	&	-0.60669	&	0.40658	\\
90.453	&	-107.435	&	-17.878	&	0.98881	&	-0.14001	&	-0.05160	\\
66.616	&	-106.297	&	-27.520	&	-0.19064	&	-0.01155	&	0.98159	\\
102.057	&	-90.266	&	14.651	&	-0.01706	&	0.30598	&	-0.95189	\\
116.331	&	-106.622	&	42.375	&	-0.68316	&	-0.34862	&	0.64168	\\
99.453	&	-104.109	&	37.249	&	0.98880	&	-0.14800	&	0.01935	\\
75.616	&	-107.582	&	28.183	&	-0.19053	&	0.44582	&	0.87461	\\
111.057	&	-73.790	&	58.073	&	-0.01708	&	-0.17142	&	-0.98505	\\
125.331	&	-75.389	&	90.223	&	-0.68311	&	-0.00997	&	0.73025	\\
108.453	&	-75.546	&	84.516	&	0.98884	&	-0.12178	&	0.08577	\\
84.616	&	-82.834	&	78.102	&	-0.19055	&	0.80145	&	0.56689	\\
120.057	&	-39.023	&	88.865	&	-0.01703	&	-0.60939	&	-0.79269	\\
134.331	&	-25.497	&	118.074	&	-0.68314	&	0.33005	&	0.65145	\\
117.453	&	-28.288	&	113.094	&	0.98888	&	-0.06776	&	0.13236	\\
93.616	&	-37.723	&	110.802	&	-0.19064	&	0.97306	&	0.12964	\\
129.057	&	6.073	&	99.971	&	-0.01707	&	-0.90805	&	-0.41851	\\
143.331	&	31.624	&	119.550	&	-0.68313	&	0.59519	&	0.42318	\\
126.453	&	26.838	&	116.437	&	0.98891	&	0.00153	&	0.14853	\\
102.616	&	17.419	&	118.792	&	-0.19054	&	0.92188	&	-0.33738	\\
\br
\end{tabular}
\end{indented}
\end{table}

\appendix
\begin{table}
\setcounter{section}{1}
\caption{First Spiral Dipole Positions (\AA) and Unit Vectors. \textit{(cont.)}}
\lineup
\begin{indented}
\item[]\begin{tabular}{@{}llllll}
\br
$x$	&	$y$	&	$z$	&	$\hat{\mu}_x$	&	$\hat{\mu}_y$	&	$\hat{\mu}_z$	\\
\mr
138.057	&	51.164	&	88.850	&	-0.01732	&	-0.99853	&	0.05131	\\
152.331	&	82.887	&	94.311	&	-0.68312	&	0.72371	&	0.09798	\\
135.453	&	77.203	&	93.780	&	0.98888	&	0.07031	&	0.13106	\\
111.616	&	69.956	&	100.242	&	-0.19074	&	0.65933	&	-0.72726	\\
147.057	&	85.922	&	58.046	&	-0.01708	&	-0.86020	&	0.50968	\\
161.331	&	116.549	&	48.140	&	-0.68314	&	0.68644	&	-0.24926	\\
144.453	&	111.269	&	50.311	&	0.98885	&	0.12336	&	0.08342	\\
120.616	&	107.856	&	59.401	&	-0.19053	&	0.24575	&	-0.95042	\\
\br
\end{tabular}
\end{indented}
\end{table}

\newpage

\section{Comparison between dipole-dipole and radiative Hamiltonians}
\label{app-b}

Here we compare the dipole-dipole Hamiltonian (Dipole model), which includes only the Hermitian
part of the coupling in Eq.~(\ref{real}), with the full radiative non-Hermitian Hamiltonian given in Eqs.~(\ref{Hmuk}-\ref{eq:g1}) (nH model). We will also include in our analysis the Hermitian part of the full radiative non-Hermitian Hamiltonian (H model). 
For the three models (Dipole, nH, and H) we compare both the real-valued energies and the dipole coupling strengths of their eigenstates. We will show that, in the small volume limit $L/\lambda \ll 1$, both quantities can be computed with the three models, but when the system size is larger than the wavelength, 
only the nH model can be used to compute the dipole strengths of the eigenstates. However, the H model still gives a close estimation to the nH model for the real energies in the large-volume limit, though the Dipole model displays deviations from the nH model values. 

\begin{figure}[ht!]
    \centering
\includegraphics[ trim=1cm 0 12 0,scale=0.5]{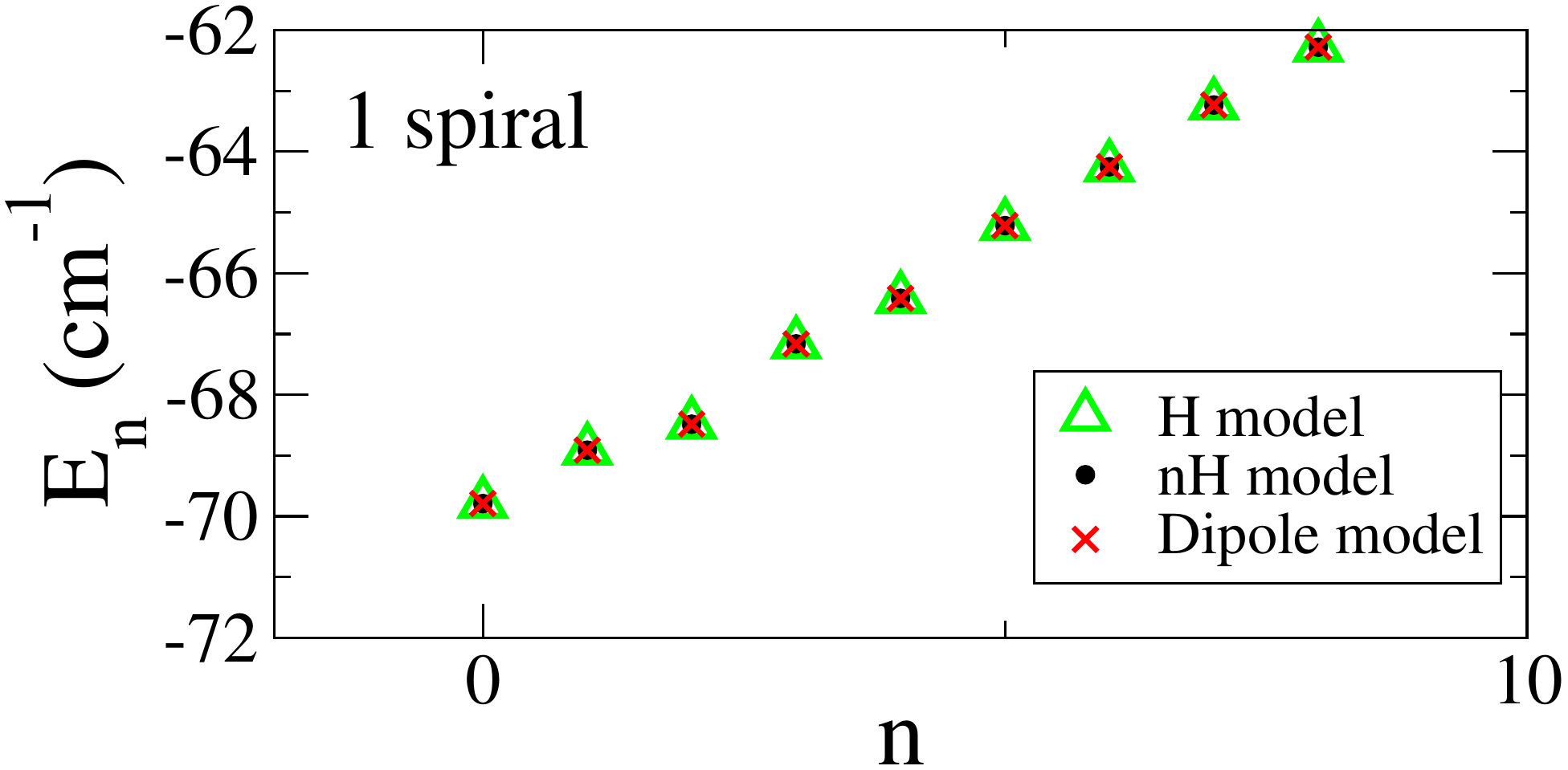}
\includegraphics[ trim=1cm 0 12 0,scale=0.5]{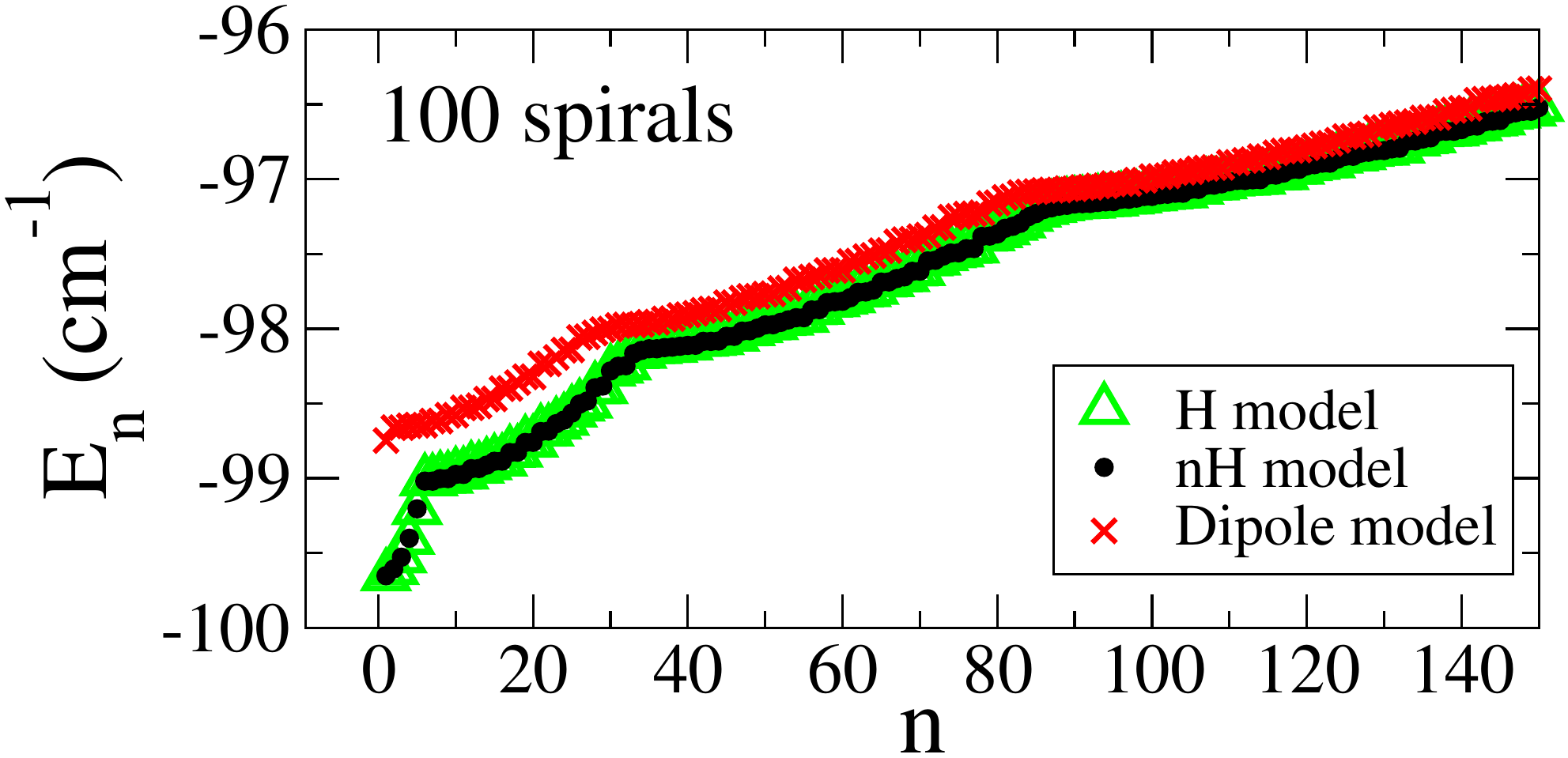}
\caption{Lowest part of the spectrum  (real-valued energies $E_n$ vs. eigenstate index $n$) for a microtubule of 1 spiral, $L/\lambda \ll 1$, (upper panel) and 100 spirals, $L/\lambda \approx 3$, (lower panel) is shown for the  three different models considered (see main text). In the small-volume limit (upper panel), all three models give similar estimations of the spectrum, but in the large-volume limit (lower panel) the Dipole model deviates from the H and nH models, which are very close to each other.
} 
\label{f1b}
\end{figure}

In  Figure~\ref{f1b}, we compare the real part of the spectrum for the three models, focusing on the eigenvalues close to the ground state. 
In the upper panel, we present a microtubule made of only one spiral, so that $L/\lambda \ll 1$. In this case the three models all give very similar estimations of the eigenvalues. In the lower panel, the case of a microtubule of 100 spirals is considered. In this case the system size is not small compared with the wavelength, as $L/\lambda \approx 3$.
One can see that while the H model is a very good approximation of the nH model, the Dipole model exhibits maximum deviations of $\sim1$ cm$^{-1}$ at and near the ground state.

\begin{figure}[ht!]
    \centering
\includegraphics[ trim=1cm 0 12 0,scale=0.5]{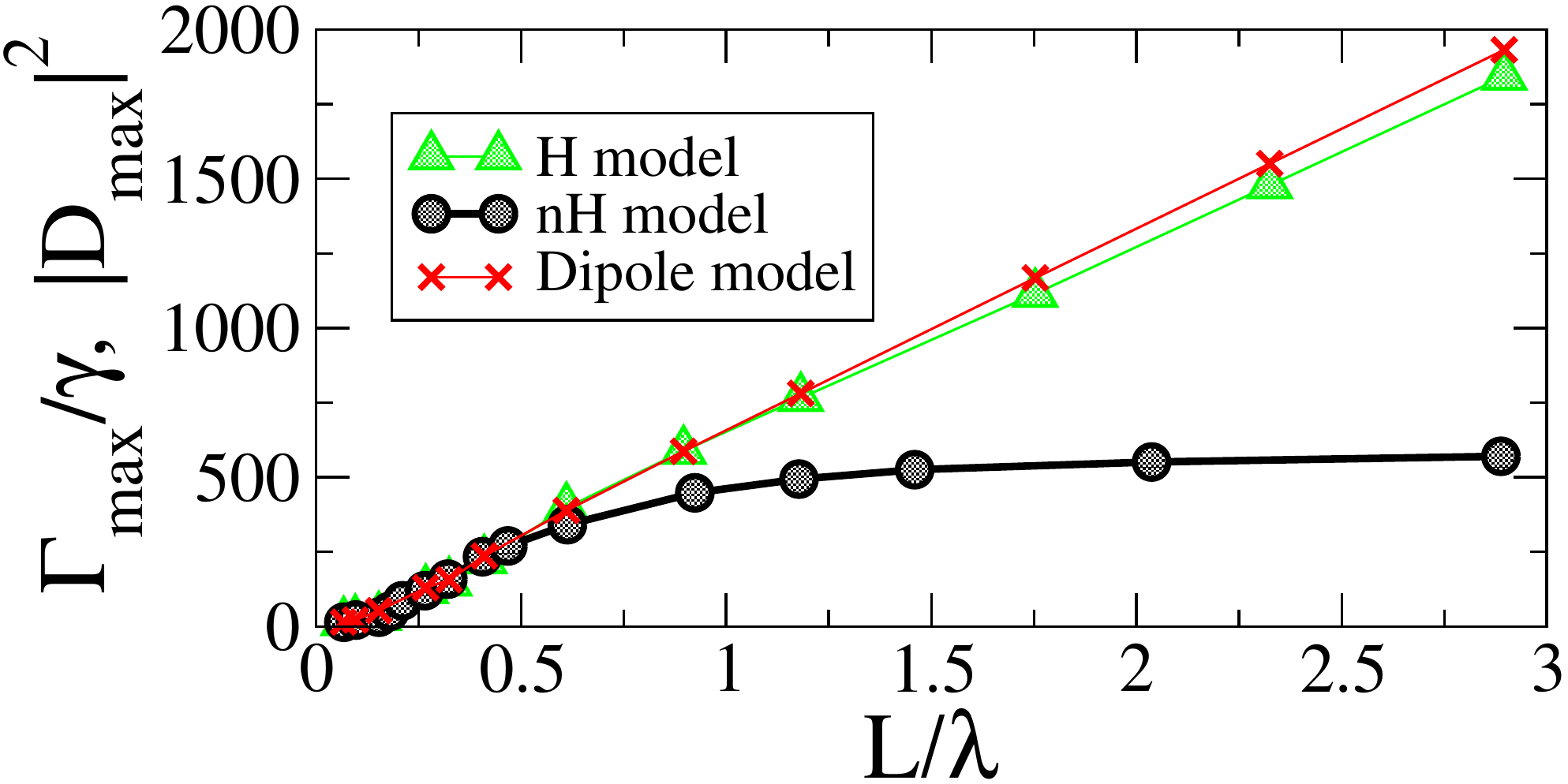}
\caption{Maximum dipole coupling strength $|D_{max}|^2$ computed from Eq.~(\ref{eq:dipst2}) for the Dipole and H models (see main text) is compared with the relative decay width $\Gamma_{max}/\gamma$ computed from the full radiative nH model (see main text) as a function of the system size $L$ normalized by the excitation wavelength ($\lambda=280$ nm).
} 
\label{f2b}
\end{figure}

When the system size is small compared to the wavelength associated with the optical transition of the molecules,
the optical absorption of an eigenstate of the aggregate can be  estimated in terms of its dipole strength, computed only from the Hermitian part of the Hamiltonian (\ref{Hmuk}). 
Denoting the $n^{th}$ eigenstate of the Hermitian part of the Hamiltonian (\ref{Hmuk}) or of the Hamiltonian with only Hermitian coupling in (\ref{real}) as $|E_n\rangle$, we can expand it in the site basis, so that 
\begin{equation} \label{eq:expan}
|E_{n}\rangle=\sum_{i=1}^{N} C_{ni} \, |i\rangle.
\end{equation}
Note that the site basis is referred to by the tryptophan molecules and is composed of the states $|i\rangle$, each of them carrying a dipole moment $\vec{\mu}_i$.  
If $N$ is the total number of molecules, then we will express the transition   dipole moment $\vec{D}_n$ associated with the $n^{th}$ eigenstate as follows: 
\begin{equation} \label{eq:dipst2} 
\vec{D}_n=\sum_{i=1}^{N} C_{ni} \, \hat{\mu}_i. 
\end{equation} 
The dipole coupling strength (often referred to as simply the dipole strength) of the $n^{th}$ eigenstate is defined  by $|\vec{D}_n|^{2}$ (note that due to normalization $\sum_{n=1}^{N} |\vec{D}_n|^{2}=N$).  Under the approximation that  $L/\lambda \ll 1$ we have $|\vec{D}_n|^2 \approx \Gamma_n/\gamma$, where $\Gamma_n$ is given by the imaginary part of the complex eigenvalues ${\cal E}_n=E_n-i\Gamma_n/2$ of the nH model. On the other hand, in the large-volume limit, the dipole as defined above in Eq.~(\ref{eq:dipst2}) gives incorrect results and does not represent the dipole of the eigenstates. This is shown in Figure~\ref{f2b}, where the maximum dipole strength computed using the Dipole model and the H model is compared with the maximum decay width $\Gamma_{\rm max}/\gamma$ computed with the full radiative nH model. As one can see, the dipole coupling strength computed as described above is valid only for small system sizes.

\section{Supertransfer and the energy gap in the complex plane}
\label{app-gap}

\begin{figure}[ht!]
    \centering
\includegraphics[ trim=1cm 0 12 0,scale=0.4]{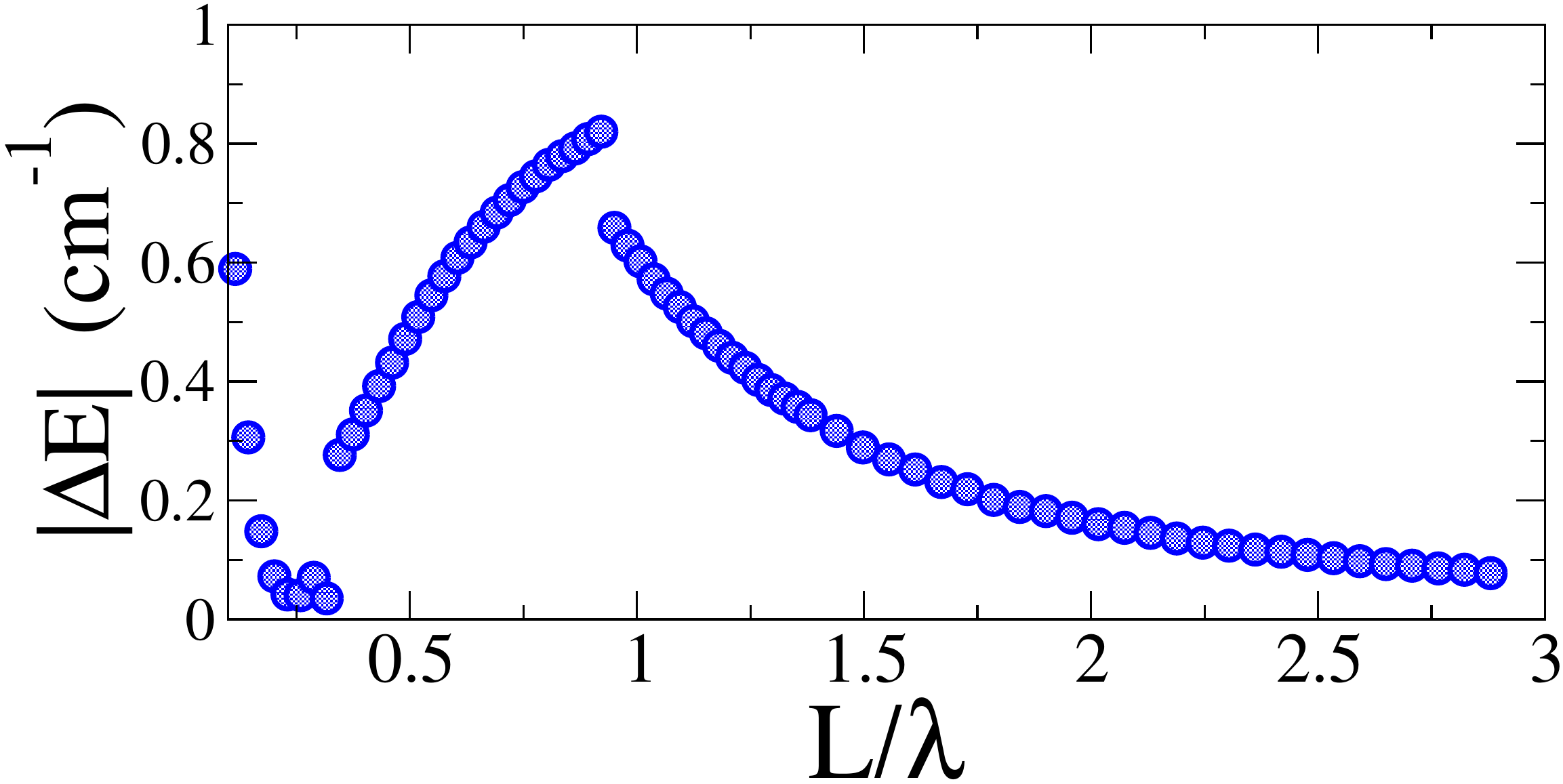}
\caption{Energy gaps in the complex plane between the ground state and the first excited state. The gap is plotted as a function of the length of the microtubule segment normalized by the wavelength of the excitation energy of the tryptophan chromophores ($\lambda=280$ nm).
} 
\label{gap}
\end{figure}

We would like to point out that supertransfer might also play an important role in stimulating robustness to disorder. For instance, in Fig.~(\ref{gap}) we show the energy differences in the complex plane between the ground state (which coincides with the most superradiant state for a microtubule of more than 12 spirals) and the first excited state. As one can see, the energy gap increases with the system size, instead of decreasing as one would expect, for lengths up to the excitation wavelength. Such counterintuitive behavior for the energy gap has analogously been found in photosynthetic complexes by two of the authors of this paper~\cite{gulli}, where it has been connected to the presence of supertransfer. It is well known that such energy gaps can protect states from disorder, but the precise consequences for robustness of this gap in cylindrical aggregates need to be studied more carefully. We plan to do this in the future. 
 
\clearpage
\section*{References} 


\begin{thebibliography}{99}
\bibitem{dipsquare} Let us recall that the units of the dipole-dipole interaction energy are given by $[E] = [\mu]^2 [d]^{-3}$, where $[\mu]$ is the dipole unit and $[d]$ the distance unit. We express the dipoles in D (Debye), the distance in \AA, and the energy in $\text{cm}^{-1}$ units (applying the standard conversion $E/(hc)$, with $h$ being the Planck constant and $c$ the speed of light), so that $[\mu^2/(hc)]=\text{cm}^{-1}\text{\AA}^3$. Now, from the definition $1 \mbox{ D}= 10^{-18} \mbox{ cm}^{5/2}\mbox{ g}^{1/2}\mbox{ s}^{-1}$ we have $1 \mbox{ D}^2= 10^{-12} \mbox{ cm}^2\mbox{ g}\mbox{ s}^{-2}\mbox{ \AA}^{3}$. Recalling the Planck constant $h = 6.626\cdot 10^{-27}\text{ cm}^2\text{ g s}^{-1}$ and the speed of light $c = 2.998\cdot 10^{10}\text{ cm s}^{-1}$, we have $1\text{ D}^2/(hc) = 5034\text{ cm}^{-1}\text{ \AA}^3$. So, a transition dipole $\mu = \sqrt{36}$~D results in $\mu^2 = 36 \times 5034\text{ cm}^{-1}\text{ \AA}^3 = 181224\text{ cm}^{-1}\text{ \AA}^3$. Note that in these calculations we write explicitly where the energy is divided by $hc$ for clarity, while in the main text we always assume implicitly that any energy is divided by $hc$.
\end{thebibliography}
\end{document}